\documentclass[12pt, draftclsnofoot, onecolumn]{IEEEtran}
\usepackage[cmex10]{amsmath}
\usepackage{graphicx}
\usepackage{algorithm}
\usepackage{algorithmic}
\usepackage{stfloats}
\usepackage{epstopdf}
\usepackage[numbers,sort&compress]{natbib}
\usepackage{multirow}
\usepackage{color}
\usepackage{balance}
\usepackage{amsmath, amssymb}
\usepackage{caption}
\usepackage{subfigure}
\usepackage{hyperref}

\usepackage{bm}
\ifCLASSINFOpdf
\else
\fi
\begin{document}
\title{Intelligent Omni Surface-Assisted Self-Interference Cancellation for Full-Duplex MISO System}
\author{Sisai~Fang,~\IEEEmembership{Student Member,~IEEE,}
Gaojie~Chen,~\IEEEmembership{Senior Member,~IEEE,}
Pei~Xiao,~\IEEEmembership{Senior Member,~IEEE,}
Kai-Kit~Wong,~\IEEEmembership{Fellow,~IEEE,}\\
Rahim Tafazolli,~\IEEEmembership{Senior Member,~IEEE}
\thanks{\noindent Sisai Fang is with the School of Engineering, University of Leicester, Leicester LE1 7RH, U.K. (e-mail: sf305@leicester.ac.uk).}
\thanks{Gaojie Chen is with the Institute for Communication Systems (ICS), 5GIC \& 6GIC, University of Surrey, Guildford, Surrey GU2 7XH, U.K. and also with the School of Engineering, University of Leicester, Leicester LE1 7RH, U.K. (e-mail: gaojie.chen@surrey.ac.uk). }
\thanks{Pei Xiao and Rahim Tafazolli are with the Institute for Communication Systems (ICS), 5GIC \& 6GIC, University of Surrey, Guildford, Surrey GU2 7XH, U.K. (e-mail: p.xiao@surrey.ac.uk and r.tafazolli@surrey.ac.uk).}
\thanks{Kai-Kit Wong is with the Department of Electronic and Electrical Engineering, University College London, London WC1E 7JE, U.K. (e-mail:
kai-kit.wong@ucl.ac.uk).}
}
\maketitle

\vspace{-2em}
\begin{abstract}
The full-duplex (FD) communication can achieve higher spectrum efficiency than conventional half-duplex (HD) communication; however, self-interference (SI) is the key hurdle. This paper is the first work to propose the intelligent Omni surface (IOS)-assisted FD multi-input single-output (MISO) FD communication systems to mitigate SI, which solves the frequency-selectivity issue. In particular, two types of IOS are proposed, energy splitting (ES)-IOS and mode switching (MS)-IOS. We aim to maximize data rate and minimize SI power by optimizing the beamforming vectors, amplitudes and phase shifts for the ES-IOS and the mode selection and phase shifts for the MS-IOS. However, the formulated problems are non-convex and challenging to tackle directly. Thus, we design alternative optimization algorithms to solve the problems iteratively. Specifically, the quadratic constraint quadratic programming (QCQP) is employed for the beamforming optimizations, amplitudes and phase shifts optimizations for the ES-IOS and phase shifts optimizations for the MS-IOS. Nevertheless, the binary variables of the MS-IOS render the mode selection optimization intractable, and then we resort to semidefinite relaxation (SDR) and Gaussian randomization procedure to solve it. Simulation results validate the proposed algorithms' efficacy and show the effectiveness of both the IOSs in mitigating SI compared to the case without an IOS.
\end{abstract}

\begin{IEEEkeywords}
Full-duplex communication, self-interference cancellation, intelligent Omni surface, MISO communication, semidenifite relaxation, non-convex optimization.
\end{IEEEkeywords}

\IEEEpeerreviewmaketitle

\section{Introduction}
The full-duplex (FD) communication has been developed to meet the ever-increasing traffic demand for wireless communications. Unlike the conventional half-duplex (HD) communication, FD communication can ideally double the ergodic capacity since it can transmit and receive in the same time slots and frequency band \cite{Challenges,self-interference}. It not only improves the ergodic capacity, but also enhances network secrecy \cite{gaojiepls} and reduces signaling overhead, and end-to-end delay \cite{Survey}. Thanks to these features, FD communication has gained much attention from industry and academia. In \cite{Allocation}, the authors applied multiple FD nodes to maximize the sum-rate of the nodes by jointly optimizing the subcarrier assignment and power allocation.
Also, a novel two-phase protocol was proposed in a wireless communication FD network which ensured uninterrupted information transmissions and self-energy recycling by exploiting the features of FD nodes \cite{Powered}. An FD receiver was introduced in the mmWave communication network \cite{covert} to cover the presence of a transmitter in the presence of a watchful warden by jointly optimizing the beamforming, transmit power and jamming. Besides, the authors of \cite{D2D} investigated an FD relay in device-to-device communications underlaying a cellular network from both analysis and optimization perspectives. To boost the sum-throughput and reduce the total-energy of Internet-of-Things networks, a rotary-wing unmanned aerial vehicle equipped with an FD access point was introduced \cite{Rotary}. In addition, the authors of \cite{BufferAided} utilized a double deep Q-network to solve the throughput maximization and secrecy rate maximization problems in a secure cognitive radio relay network where the relay nodes operated in the FD mode and transmitted jamming signals to the eavesdropper. Furthermore, an optimal power allocation strategy was proposed for the dual-hop FD decode-and-forward relay system to minimize the outage probability. However, the FD system cannot achieve the ideal performance due to the self-interference (SI) \cite{Dual}.

As a downside of FD communications, SI imposed by the devices' own transmissions cannot be neglected. The SI, if left unattended, could result in reduced capacity for FD systems that fall below that of HD systems \cite{self-interference}.   Fortunately, self-interference cancellation (SIC) techniques have been exploited to tackle this issue. The SIC techniques can be classified into three steps: passive suppression, analogue cancellation and digital cancellation. The passive suppression relies on a large physical distance between transmit and receive antennas, thus high isolation can be achieved \cite{Applications}. Secondly, the analogue cancellation exploits the degrees of freedom bestowed by antenna arrays. Specifically, the beam patterns of transmitters can be designed to suppress the signal strength received at the receive antennas \cite{Duplexing, BeamBased}.  Furthermore, the digital cancellation technique applies digital cancellation protocols, such as ZigZag \cite{Jigsaw} to mitigate residual SI further. The authors of \cite{AllDigital} proposed a novel digital SIC technique to mitigate the SI signal and the transmitter and receiver impairments. 
However, conventional RF cancellation techniques are frequency-selective due to analogue SI cancellation circuit, such as Balun, i.e., the best cancellation result can be achieved at the center frequency, and the SIC result is not promising for frequency components that are far apart from the center frequency \cite{R1}. This imposes severe bandwidth limitations for FD radio, which would fail to operate at the high-frequency bands such as millimetre waves. Other problems include implementation complexity and large insertion loss associated with RF SIC circuits. In addition, it was demonstrated that the operational bandwidth and the performance of RF-based cancellation were usually restricted by the hardware \cite{wideband}. Moreover, although SI can be reduced via designing the beam pattern of the transmitter \cite{largeantenna}, the size, weight, and power (SWaP) limitations of small devices do not allow the deployment of enough RF units to achieve flexible and efficient FD transmissions \cite{Non_Terrestrial}. In a nutshell, the SIC is still a key hurdle to FD communications to double the channel capacity for wireless communication networks.

In recent years, the reconfigurable intelligent surface (RIS) has been exploited to meet the increasing demands for future sixth-generation (6G) wireless networks because it can enhance the performance of wireless communication at low costs. In principle, RIS is constructed by a number of components whose electromagnetic responses can be adjusted to enhance a given performance metric \cite{Cognitive}. The research regarding RIS has been spreading in various areas such as  unmanned aerial vehicle communications \cite{Sisai}, cooperative networks \cite{chongRIS}, dual-function radar communication systems \cite{GC}, physical layer security enhancement \cite{magRIS}. In particular, RIS supports wide frequency bands such as mmWave, the authors of \cite{RISmmW} demonstrated that the RIS was capable of improving the sum-rate of the proposed mmWave systems.

Moreover, it has been demonstrated that the utility of RISs in FD communications can enhance the performance of wireless communication networks \cite{Zaid, Robust}. For example, the weighted sum-rate of the multi-input multi-output (MISO) communication system with an FD base station can be improved by introducing a RIS \cite{Unfolding}.  Additionally, the weighted sum transmit power consumption of the base station and uplink user can be greatly reduced by jointly optimizing their transmit power, and the phase shifts \cite{Aided}, subject to the rate constraints at the users and uni-modulus constraints at the RISs.  Furthermore, the authors studied the RIS technology in an FD wireless communication system where the outage and error probabilities were derived \cite{Achieve}. More specifically, the authors provided some insights into how to enhance the performance and save energy consumption, via obtaining the closed-form expressions of ergodic capacity and symbol error rate, in the RIS assisted FD systems \cite{Impairments}. Besides, the authors of \cite{TwoWay} investigated a RIS-aided FD wireless communication system and demonstrated that the RIS was capable of enhancing the performance of FD systems with a large number of elements when the SI cannot be ideally removed. Notice that the above works only use the RISs to enhance signal strength at the receiver, and they have not utilized the RIS to balance the SIC and data transmission.  In addition, all of the above works assumed that the traditional SIC techniques could eliminate the SI instead of mitigating the SI with the assistance of RISs. Therefore, the aforementioned frequency-selectivity and SWaP problems for the conventional SIC schemes have not yet been well addressed.

To tackle this problem,  we utilize the intelligent Omni surface (IOS), which was developed by NTT DOCOMO INC. to achieve reflection and refraction in 360-degree coverage \cite{NTT}, in reducing SI as well as boosting the transmission rate of FD communication networks, and solving the bandwidth constraint and SWaP limitation incurred by traditional SIC techniques. To the best of our knowledge, this work is the first of its kind to establish an IOS-aided FD MISO wireless communication network in cancelling SI and enhancing data rate. Specifically, we consider two modes of the IOS, namely, energy splitting (ES)-IOS and mode selection (MS)-IOS, to maximize the data rate while minimizing the SI power, subject to the data rate threshold. The main contributions of this paper are summarized as follows:

\begin{enumerate}
\item We propose a novel IOS-aided MISO FD communication network and introduce two types of IOS, ES-IOS and MS-IOS. The proposed schemes not only can enhance the data rate at the destination but also cancel the SI and against the frequency-selectivity and SWaP limitations issues that exist widely in the traditional SIC schemes.
\item We investigate two optimization problems, including data rate maximization and SI power minimization problems. Firstly, we aim to maximize the data rate at the destination, subject to the transmit power, SI power, and phase shifts constraints. On the other hand, SI power is minimized, subject to the data rate threshold, transmit power and phase shifts constraints.
\item We utilize alternative algorithms for two non-convex problems with different types of the IOS to solve the sub-problems iteratively. The quadratically constrained quadratic programming (QCQP)  method is applied to solve the beamforming, joint amplitudes and phase shifts optimizations with the ES-IOS  as well as the phase shifts optimizations with the MS-IOS . Besides, we resort to the semi-definite relaxation (SDR) and Gaussian randomization method to solve the binary optimizations
for the cases with MS-IOS,
\item Simulation results validate the efficiency of the proposed algorithms and demonstrate the flexibility and effectiveness of utility of IOS to strike a good balance  between SIC and data transmission for FD MISO communication networks. More specifically, we show that the SI can be reduced to the noise floor. At the same time, the data rate requirement is guaranteed for ES-IOS and MS-IOS, which shows the effective implementation of IOS in MISO FD communication networks.
\end{enumerate}

The remainder of this paper is organized as follows. Section II presents the system model of our proposed IOS-assisted MISO FD communication system and the problem formulations. Section III maximizes the data rate at the destination by optimizing the beamforming vectors of the transmitter, the amplitudes and phase shifts of the ES-IOS, the mode selection and phase shifts of the MS-IOS, iteratively. Section IV presents the SI power minimization for the cases with ES-IOS and MS-IOS, respectively. Section V presents numerical results to validate the efficiency of our proposed algorithms and the merits of applying IOS to the MISO FD communication system. Finally, Section VI concludes this paper.

\textit{Notations:} In this paper, scalars are denoted by italic letters, vectors and matrices are denoted by bold lowercase letters and bold uppercase letters letters, respectively. $\mathbf{a}^H$ gives the Hermitian of the vector $\mathbf{a}$, $\mathbf{a}^T$ and $\mathbf{a}^*$ denote its transpose and conjugate operators, respectively. $\left[\mathbf{A} \right]_{i, j}$ is the element located on the $i$th row and $j$th column of matrix $\mathbf{A}$, ${\left[ {\mathbf{a}} \right]_i}$ is the $i$th element of vector ${\mathbf{a}}$. ${\mathbf{B}} \odot {{\mathbf{C}}^T}$ is the Hadamard product of ${\mathbf{B}}$ and ${\mathbf{C}}$. The trace of a matrix $\mathbf{A}$ is represented by $\text{Tr}(\mathbf{A})$. $\operatorname{Re} \left\{  \cdot  \right\}$ denotes the real part of a complex value,
${\left\| {\mathbf{A}} \right\|_F}$ represents the Frobenius norm of matrix $\mathbf{A}$,
$\textbf{1}_M$ stands for the $M\times 1$ identity vector.  ${\text{diag}}\left\{  \cdot  \right\}$ and ${\left(  \cdot  \right)^*}$ denote the operator for diagonalization and the optimal value, respectively and $\mathcal{O}\left(  \cdot  \right)$ is the big-O notation.

\section{System model and problem formulation}

\subsection{System model}

\begin{figure*}[h]
	\centering
	\subfigure[]{
		\begin{minipage}[b]{0.45\textwidth}
			\includegraphics[width=1\textwidth]{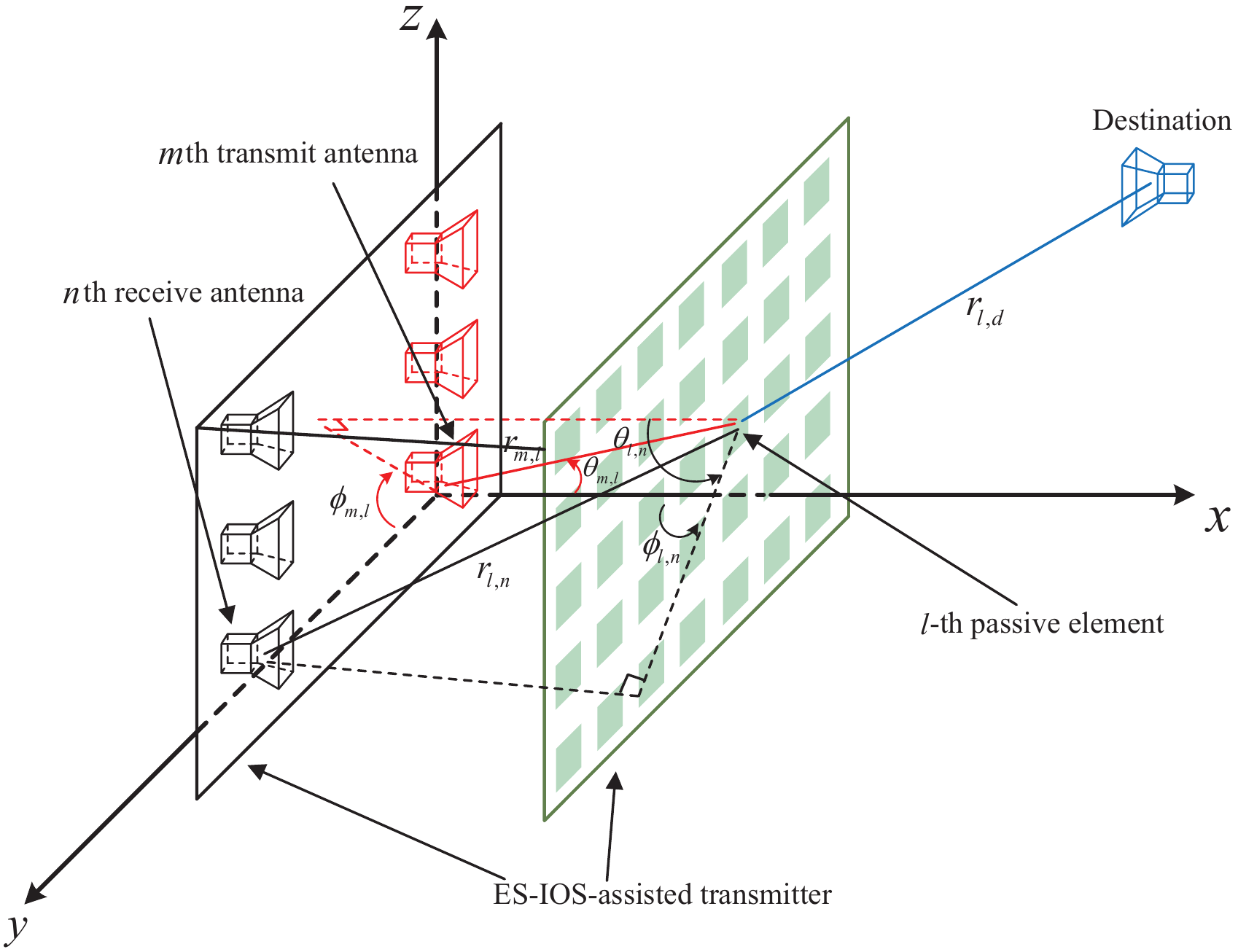}
		\end{minipage}
		\label{3D}
	}
    	\subfigure[]{
    		\begin{minipage}[b]{0.45\textwidth}
   		 	\includegraphics[width=1\textwidth]{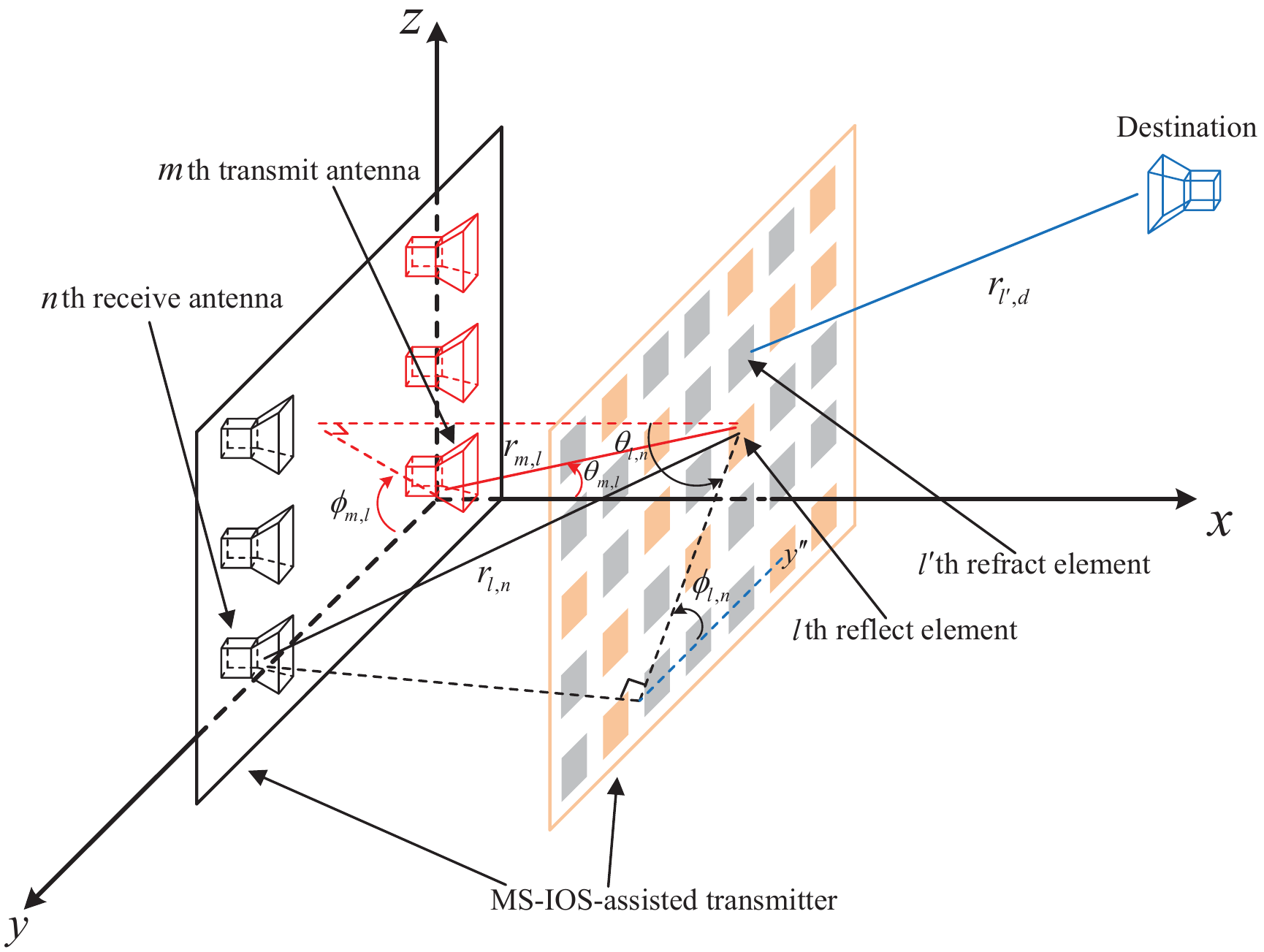}
    		\end{minipage}
		\label{3D_2}
    	}
	\caption{(a) An ES-IOS assisted MISO FD network.  (b) An MS-IOS assisted MISO FD network. }
	\label{fig:hor_2figs_1cap_2subcap}
\end{figure*}

We design an IOS-assisted FD MISO communication network where the transmitter sends signals to the single-antenna destination with the assistance of a $L$-element IOS; and simultaneously design the phase shifts of the IOS to mitigate the SI between the receive antennas and the transmit antennas. The transmitter is equipped with $M$ transmit antennas and $N$ receive antennas and integrated with an IOS. Furthermore, two types of IOSs are employed: 1) ES-IOS, the elements capable of reflecting and refracting signals simultaneously, as illustrated in Fig. \ref{3D}.  2) MS-IOS\footnote{It is worth mentioning that the MS-IOS is more readily implementable than the ES-IOS because of the low hardware design complexity, while the ES-IOS provides higher flexibility \cite{implement}; therefore, in this paper, we investigate both cases.}, where each element can only reflect or refract signal at one-time slot, is illustrated in Fig. \ref{3D_2}.  In addition, the reflecting and refracting phase shifts coefficient matrices of the ES-IOS are represented as ${{\mathbf{\Theta }}_e} = {\text{diag}}\left\{ {{a_{e,1}}{e^{j{\alpha _{e,1}}}},{a_{e,2}}{e^{j{\alpha _{e,2}}}}, \ldots ,{a_{e,L}}{e^{j{\alpha _{e,L}}}}} \right\} \in {\mathbb{C}^{L \times L}}$ and ${{\mathbf{\Phi }}_e} \!=\!{\text{diag}}\left\{ {{b_{e,1}}{e^{j{\beta _{e,1}}}},{b_{e,2}}{e^{j{\beta _{e,2}}}}, \ldots ,{b_{e,L}}{e^{j{\beta _{e,L}}}}} \right\} \!\in\! {\mathbb{C}^{L \times L}}$, where $a_{e,l}$, $b_{e,l}$ denote the reflecting and refracting amplitudes at the $l$th element, $l = 1,2, \ldots , L$, and $\alpha_{e,l}$, $\beta_{e,l}$ indicate the phase shifts of the $l$th element, respectively. Hence, according to \cite{15}, the constraints \footnote{Note that continuous phase shifts are considered in this paper, similar to \cite{Robust,Unfolding,Aided}, which is the upper bound of the system performance. Besides, the discrete phase shifts can be derived by quantifying the obtained optimal continuous phase shifts, which will be shown in Section V.} for the ES-IOS can be given by
\begin{subequations}\label{constraint_1}
\begin{equation}\small\label{1a}
a_{e,l}^2 + b_{e,l}^2 \leq 1,
\end{equation}
\begin{equation}\small\label{1b}
0 \leq{a_{e,l}},{b_{e,l}} \leq 1,
\end{equation}
\begin{equation}\small\label{1c}
0 \leq {\alpha _{e,l}},{\beta _{e,l}} \leq 2\pi , \ \  \forall l.
\end{equation}
\end{subequations}
On the other hand, the reflecting and refracting phase shifts coefficient matrices of the MS-IOS are denoted as ${{\mathbf{\Theta }}_m} = {\text{diag}}\left\{ {{a_{m,1}}{e^{j{\alpha _{m,1}}}},{a_{m,2}}{e^{j{\alpha _{m,2}}}}, \ldots ,{a_{m,L}}{e^{j{\alpha _{m,L}}}}} \right\} \in {\mathbb{C}^{L \times L}}$ and ${{\mathbf{\Phi }}_m} = {\text{diag}}\left\{ {{b_{m,1}}{e^{j{\beta _{m,1}}}},{b_{m,2}}{e^{j{\beta _{m,2}}}}, \ldots ,{b_{m,L}}{e^{j{\beta _{m,L}}}}} \right\} \in {\mathbb{C}^{L \times L}}$
and the constraints for the MS-IOS are listed as follows:
\begin{subequations}\label{constraint_2}
\begin{equation}\small\label{2a}
{a_{m,l}} + {b_{m,l}} = 1,
\end{equation}
\begin{equation}\small\label{2b}
{a_{m,l}},{b_{m,l}} \in \left\{ {0,1} \right\},
\end{equation}
\begin{equation}\small\label{2c}
0 \leq {\alpha _{m,l}},{\beta _{m,l}} \leq 2\pi,  \ \  \forall l,
\end{equation}
\end{subequations}
where \eqref{2a} indicates that the $l$th element of the MS-IOS operates in either reflection or refraction mode. The transmit signal at the transmitter is given by
\begin{equation}\small
{\mathbf{x}} = {\mathbf{w}s},
\end{equation}
where $\mathbf{w}\in {\mathbb{C}^{M \times 1}}$ is the baseband beamforming vector, $s$ is the  transmit signal. Besides, we denote the channel coefficients between the transmit antennas and the IOS, the transmit antennas and the receive antennas, the IOS (ES-IOS/MS-IOS) and the destination, the IOS and the receive antennas as ${{\mathbf{H}}_{ti}} \in {\mathbb{C}^{L \times M}}$, ${{\mathbf{H}}_{tr}} \in {\mathbb{C}^{N \times M}}$, ${{\mathbf{h}}_{id}} \in {\mathbb{C}^{L \times 1}}$ and ${{\mathbf{H}}_{ir}} \in {\mathbb{C}^{L \times N}}$, respectively.
Therefore,  the received signals \footnote{We assume that the ES/MS-IOS-assisted transmitter has the knowledge of the channel state information (CSI) of the transmitter-IOS-destination link, as well as the link between transmit antennas and receive antennas. The details of the channel estimation can be found in \cite{CSI1, CSI2}, which is out of the scope of this paper. Besides, to provide the further information, we compare the imperfect CSI case as a benchmark in Section V.}
at the destination and the receiver with ES-IOS and MS-IOS are given by
\begin{equation}\label{signal_d}\small
{{\mathbf{y}}_{d,x}} = {\mathbf{h}}_{id}^H{{\mathbf{\Phi }}_x}{{\mathbf{H}}_{ti}}{\mathbf{x}} + {{\mathbf{n}}_d},
\end{equation}
\begin{equation}\label{signal_r}\small
{{\mathbf{y}}_{r,x}} = \left( {{\mathbf{H}}_{tr}^H + {\mathbf{H}}_{ir}^H{{\mathbf{\Theta }}_x}{{\mathbf{H}}_{ti}}} \right){\mathbf{x}} + {{\mathbf{n}}_r},
\end{equation}
respectively, where $x = \left\{ {e,m} \right\}$ denotes the case with ES-IOS and MS-IOS. ${\mathbf{n}_d} \sim \mathcal{CN}\left( {0,\sigma _d^2{\mathbf{I}}} \right)$ and ${\mathbf{n}_r} \sim \mathcal{CN}\left( {0,\sigma _r^2{\mathbf{I}}} \right)$ represent the additive Gaussian noise received at the destination and receive antennas, respectively.  $\sigma _d^2$ and $\sigma _r^2$ are the noise powers at the destination and the receiver antennas, respectively.  As illustrated in Fig. \ref{3D} and Fig. \ref{3D_2}, to facilitate presentation, we characterize the positions of the transmit antennas by $\left( {{r_{l,m}},\theta _{l,m},\phi _{l,m}} \right)$, ${r_{l,m}} \geq 0$, $\theta _{l,m} \in \left[ {0,\frac{\pi }{2}} \right]$, $\phi _{l,m} \in \left[ {0,2\pi } \right]$, in $L$ different spherical coordinate systems whose respective origins are the positions of the IOS elements \cite{FDmainref}. Here, ${\theta _{l,m}}$ denotes the elevation angle of the $m$th transmit antenna, ${\phi _{l,m}}$ represents the azimuth angle of the $m$th transmit antenna, and ${r_{l,m}}$ is the distance between the $l$th IOS element and the $m$th transmit antenna. In a similar manner, we have the positions of the receive antennas at the elements of the IOS, the positions of the receive antennas at the transmit antennas, and the positions of the transmit antennas at the receive antennas as  $\left( {{r_{l,n}},{\theta _{l,n}},{\phi _{l,n}}} \right)$,
$\left( {{r_{m,n}},{\theta _{m,n}},{\phi _{m,n}}} \right)$ and $\left( {{r_{n,m}},{\theta _{n,m}},{\phi _{n,m}}} \right)$, where ${r_{l,n}},{r_{m,n}},{r_{n,m}} \geq 0$,
${\theta _{l,n}},{\theta _{m,n}},{\theta _{n,m}} \in \left[ {0,\frac{\pi }{2}} \right]$, ${\phi _{l,n}},{\phi _{m,n}},{\phi _{n,m}} \in \left[ {0,2\pi } \right]$. Then we can define the corresponding channel coefficients as follows:
\begin{equation} \small
{{\mathbf{H}}_{ti}}\! =\! {\left[ {\frac{{\lambda\! \sqrt {{G^t}\left( {{\theta _{l,m}},{\phi _{l,m}}} \right)\!} }}{{4\pi {r_{l,m}}}}\!{e^{ - j\frac{{2\pi {r_{l,m}}}}{\lambda }}}} \right]_{l,m}},
\end{equation}
\begin{equation}\small
{{\mathbf{H}}_{tr}} = {\left[ {\frac{{\lambda \sqrt {{G^t}\left( {{\theta _{m,n}},{\phi _{m,n}}} \right){G^r}\left( {{\theta _{n,m}},{\phi _{n,m}}} \right)} }}{{4\pi r_{m,n}^{{\kappa  \mathord{\left/
 {\vphantom {\kappa  2}} \right.
 \kern-\nulldelimiterspace} 2}}}}\left( {\sqrt {\frac{K}{{K + 1}}} {e^{ - j\frac{{2\pi {r_{m,n}}}}{\lambda }}} + \sqrt {\frac{1}{{K + 1}}} h_{tr}^{nlos}} \right)} \right]_{m,n}},
\end{equation}
\begin{equation}\small
{{\mathbf{h}}_{id}} \!=\! {\left[ {\frac{{\lambda }}{{4\pi r_{l,d}^{{\kappa  \mathord{\left/
 {\vphantom {\kappa  2}} \right.
 \kern-\nulldelimiterspace} 2}}}}\left( {\sqrt {\frac{K}{{K\!\! +\!\! 1}}} {e^{ -\! j\frac{{2\pi {r_{l,d}}}}{\lambda }}} \!+\! \sqrt {\frac{1}{{K \!+\!\! 1}}} h_{id}^{nlos}} \right)} \right]_{l}},
\end{equation}
\begin{equation}\small
{{\mathbf{H}}_{ir}} = {\left[ {\frac{{\lambda \sqrt {{G^r}\left( {{\theta _{l,n}},{\phi _{l,n}}} \right)} }}{{4\pi {r_{l,n}}}}{e^{ - j\frac{{2\pi {r_{l,n}}}}{\lambda }}}} \right]_{l,n}},
\end{equation}
where $\lambda $ is the wavelength, ${G^t}\left( {\theta ,\phi } \right)$ and ${G^r}\left( {\theta ,\phi } \right)$  are the antenna gains at the transmit and receive antennas\footnote{The transmit and receive antenna gain models with respect to the elevation angles can be found in \cite{FDmainref}.}, respectively.
In addition, $K$ and $\kappa$ are rician factor and exponent of the channels, $r_{l,d}$ is the distance between the $l$th element at the IOS and destination, $h_{tr}^{nlos}$ and $h_{id}^{nlos}$ are the non line-of-sight channel components, generalized using zero-mean and unit-variance circularly symmetric complex Gaussian random variables.

Then, based on \eqref{signal_d} and \eqref{signal_r}, we can derive the data rate at the destination and SI power at the receive antennas for ES-IOS and MS-IOS as follows:
\begin{equation}\small
{R_{d,x}} = {\log _2}\left( {1 + \frac{{{{\mathbf{h}}_{d,x}}{\mathbf{w}}{{\mathbf{w}}^H}{\mathbf{h}}_{d,x}^H}}{{\sigma _d^2}}} \right),
\end{equation}
\begin{equation}\small
{P_r} = {\left\| {{{\mathbf{H}}_{r,x}}{\mathbf{w}}{{\mathbf{w}}^H}{\mathbf{H}}_{r,x}^H} \right\|_F},
\end{equation}
where ${{\mathbf{h}}_{d,x}} = {\mathbf{h}}_{id}^H{\mathbf{\Phi }_x}{{\mathbf{H}}_{ti}}$, ${{\mathbf{H}}_{r,x}} = {\mathbf{H}}_{tr}^H + {\mathbf{H}}_{ir}^H{\mathbf{\Theta }_x}{{\mathbf{H}}_{ti}}$.

\subsection{Problem formulation}
In this paper, we aim to solve two optimization problems, firstly by maximizing the data rate at the destination, subject to the power consumption constraint at the transmitter, the phase shift unit modulus constraint and the received SI power at the receive antennas below a threshold:

\vspace{-1.7em}
\begin{small}\begin{align}
\mathop {\max }\limits_{\mathbf{\Theta}_x,\mathbf{\Phi}_x,\mathbf{w}}&\ \ {R_{d,x}}  \label{P1_OF}\\
{\text{s.t.}}\ \ & \ \   {\text{Tr}}\left( {{\mathbf{w}}{{\mathbf{w}}^H}} \right) \leq {P_{\max }},  \tag{\ref{P1_OF}{a}}  \label{P1_Power} \\
& \ \  \left\| {{{\mathbf{H}}_{r,x}}{\mathbf{w}}{{\mathbf{w}}^H}{\mathbf{H}}_{r,x}^H} \right\|_F \leq {P_{th}}, \tag{\ref{P1_OF}{b}}  \label{P1_sinr}\\
& \ \  \eqref{1a}-\eqref{1c} \ \text{or} \ \eqref{2a}-\eqref{2c}, \tag{\ref{P1_OF}{c}}  \label{P1_modulus}
\end{align}\end{small}where \eqref{P1_Power} denotes the power constraint of the transmitter, $P_{\max }$ is the maximum transmit power. In addition, \eqref{P1_sinr} represents the SI threshold at the receive antennas of the transmitter is below a threshold $P_{th}$. Also, it is worth mentioning that the first item of \eqref{P1_modulus} is selected for the case with ES-IOS and the second item is chosen for the case with MS-IOS.

Secondly, the SI power is minimized at the transmitter, subject to the power consumption constraint at the transmitter, the phase shift unit modulus constraint and the data rate at the destination above a threshold. Accordingly, the two optimization problems are formulated as follows:

\vspace{-1.7em}
\begin{small}\begin{align}
 \mathop {\min }\limits_{\mathbf{\Theta}_x,\mathbf{\Phi}_x,\mathbf{w}}&\ \left\| {{{\mathbf{H}}_{r,x}}{\mathbf{w}}{{\mathbf{w}}^H}{\mathbf{H}}_{r,x}^H} \right\|_F  \label{P222_OF}\\
{\text{s.t.}}\ \ & \ \   {\text{Tr}}\left( {{\mathbf{w}}{{\mathbf{w}}^H}} \right) \leq {P_{\max }},  \tag{\ref{P222_OF}{a}}  \label{P222_Power} \\
& \ \  \ {R_{d,x}}  \geq {R_{th}}, \tag{\ref{P222_OF}{b}}  \label{P222_rate}\\
& \ \  \eqref{1a}-\eqref{1c} \ \text{or} \ \eqref{2a}-\eqref{2c}, \tag{\ref{P222_OF}{c}}  \label{P222_modulus}
\end{align}\end{small}where \eqref{P222_rate} represents that the data rate threshold at the destination above a threshold $R_{th}$, and \eqref{P222_modulus} should be selected according to ES-IOS or MS-IOS mode.

\section{Maximizing the Data rate}
Given the data rate maximization problem in \eqref{P1_OF}, we notice that the problem is non-convex and difficult to solve since the reflecting, refracting amplitudes/mode selection and phase shifts $\mathbf{\Theta}_x$, $\mathbf{\Phi}_x$ and the beamforming vector $\mathbf{w}$ are coupled. Especially, the mode selection optimization for MS-IOS is NP-hard, renders the maximization problem intractable. Therefore, we propose an alternating algorithm to solve them iteratively.

\subsection{Optimizing  beamforming vector $\mathbf{w}$ for given $\mathbf{\Theta}_x$ and $\mathbf{\Phi}_x$}
Firstly, with given reflecting and refracting amplitudes and phase shifts $\mathbf{\Theta}_x$ and $\mathbf{\Phi}_x$ for ES-IOS and MS-IOS, the beamforming vector $\mathbf{w}$ optimization problem for maximizing the data rate at the destination can be reformulated as follows:

\vspace{-1.7em}
\begin{small}\begin{align}
 \mathop {\max }\limits_{\mathbf{w}}&\ \ {\log _2}\left( {1 + \frac{{{{\mathbf{h}}_{d,x}}{\mathbf{w}}{{\mathbf{w}}^H}{\mathbf{h}}_{d,x}^H}}{{\sigma _d^2}}} \right) \label{P3_OF}\\
{\text{s.t.}}  & \ \   {\text{Tr}}\left( {{\mathbf{w}}{{\mathbf{w}}^H}} \right) \leq {P_{\max }},  \tag{\ref{P3_OF}{a}}  \label{P3_Power} \\
& \ \  \left\| {{{\mathbf{H}}_{r,x}}{\mathbf{w}}{{\mathbf{w}}^H}{\mathbf{H}}_{r,x}^H} \right\|_F \leq {P_{th}}, \tag{\ref{P3_OF}{b}}  \label{P3_sinr}
\end{align}\end{small}which is equivalent to the following  problem:

\vspace{-1.7em}
\begin{small}\begin{align}
 \mathop {\max }\limits_{\mathbf{w}}& \ \  {{\mathbf{w}}^H}{\mathbf{h}}_{d,x}^H{{\mathbf{h}}_{d,x}}{\mathbf{w}} \label{P4_OF}\\
{\text{s.t.}} & \ \   {\text{Tr}}\left( {{\mathbf{w}}{{\mathbf{w}}^H}} \right) \leq {P_{\max }},  \tag{\ref{P4_OF}{a}}  \label{P4_Power} \\
& \ \  {\text{Tr}}\left( {{{\mathbf{w}}^H}{\mathbf{H}}_{r,x}^H{{\mathbf{H}}_{r,x}}{\mathbf{w}}} \right) \leq {P_{th}}, \tag{\ref{P4_OF}{b}}  \label{P4_sinr}
\end{align}\end{small}however, this problem is still non-convex due to the non-concavity of the objective function. Thus by applying the first-order Taylor expansion to \eqref{P4_OF}, the objective function can be convexified as:
\begin{equation}\small
{{\mathbf{w}}^H}\!{\mathbf{h}}_{d,x}^H{{\mathbf{h}}_{d,x}}\!{\mathbf{w}} \!\geq\! 2\operatorname{Re} \!\left\{ {{{\mathbf{w}}^H}{\mathbf{h}}_{d,x}^H\!{{\mathbf{h}}_{d,x}}{\mathbf{\tilde w}}} \right\} \! -\! {{{\mathbf{\tilde w}}}^H}\!{\mathbf{h}}_{d,x}^H{{\mathbf{h}}_{d,x}}{\mathbf{\tilde w}},
\end{equation}
where ${{\mathbf{\tilde w}}}$ is the beamforming vector value of the last iteration. Therefore, the beamforming vector optimization problem  can be formulated as follows:

\vspace{-1.7em}
\begin{small}\begin{align}
 \mathop {\max }\limits_{\mathbf{w}}& \ \  \operatorname{Re} \left\{ {{{\mathbf{w}}^H}{\mathbf{h}}_{d,x}^H{{\mathbf{h}}_{d,x}}{\mathbf{\tilde w}}} \right\} \label{P5_OF}\\
{\text{s.t.}} & \ \   {\text{Tr}}\left( {{\mathbf{w}}{{\mathbf{w}}^H}} \right) \leq {P_{\max }},  \tag{\ref{P5_OF}{a}}  \label{P5_Power} \\
& \ \  {\text{Tr}}\left( {{{\mathbf{w}}^H}{\mathbf{H}}_{r,x}^H{{\mathbf{H}}_{r,x}}{\mathbf{w}}} \right) \leq {P_{th}}, \tag{\ref{P5_OF}{b}}  \label{P5_sinr}
\end{align}\end{small}which is a QCQP problem and can be solved by convex optimization tools such as CVX.

\subsection{Optimizing reflecting and refracting amplitudes/mode selection and phase shifts $\mathbf{\Theta}_x$ and $\mathbf{\Phi}_x$ for given $\mathbf{w}$}
Secondly, since the binary optimization in mode selection for the case with MS-IOS is NP-hard,  we classify the data rate maximization problem into the two cases for a given beamforming vector $\mathbf{w}$, according to the type of IOS.\\
\\
a) \textit{Optimizing reflecting and refracting amplitudes and phase shifts $\mathbf{\Theta}_e$ and $\mathbf{\Phi}_e$ for ES-IOS}

For a given beamforming vector $\mathbf{w}$, the data rate maximization problem with the ES-IOS can be reformulated as follows:

\vspace{-1.7em}
\begin{small}\begin{align}
 \mathop {\max }\limits_{\mathbf{\Theta}_e,\mathbf{\Phi}_e}&\ \ {\mathbf{h}}_{id}^H{\mathbf{\Phi }_e}{{\mathbf{H}}_{ti}}{\mathbf{w}}{{\mathbf{w}}^H}{\mathbf{H}}_{ti}^H{{\mathbf{\Phi }}_e^H}{{\mathbf{h}}_{id}} \label{P6_OF}\\
{\text{s.t.}} & \ \
  {\text{Tr}}\left( {\left( {{\mathbf{H}}_{tr}^H \!\! +\!\! {\mathbf{H}}_{ir}^H{\mathbf{\Theta }_e}{{\mathbf{H}}_{ti}}} \right){\mathbf{w}}{{\mathbf{w}}^H}\left( {{\mathbf{H}}_{tr}^H \!\!+\!\! {\mathbf{H}}_{ti}^H{{\mathbf{\Theta }}_e^H}{{\mathbf{H}}_{ir}}} \right)} \right)
   \leq {P_{th}},  \tag{\ref{P6_OF}{a}}  \label{P6_sinr}  \\
& \ \  a_{e,l}^2 + b_{e,l}^2 \leq 1, \tag{\ref{P6_OF}{b}}  \label{P6_coeff} \\
& \ \  0 \leq{a_{e,l}},{b_{e,l}} \leq 1, \tag{\ref{P6_OF}{c}}  \label{P6_01}  \\
& \ \  0 \leq {\alpha _{e,l}},{\beta _{e,l}} \leq 2\pi , \ \  \forall l, \tag{\ref{P6_OF}{d}}  \label{P6_phase}
\end{align}\end{small}which can be reformulated as

\vspace{-1.7em}
\begin{small}\begin{align}
 \mathop {\max }\limits_{\mathbf{\Theta}_e,\mathbf{\Phi}_e}&\ \ {\text{Tr}}\left( {{{\mathbf{\Phi }}_e^H}{{\mathbf{X}}_{e,1}}{\mathbf{\Phi }_e}{{\mathbf{Y}}_{e,1}}} \right) \label{P7_OF}\\
{\text{s.t.}} & \ \
  {\text{Tr}}\left( {{\mathbf{\Theta }}_s^H{{\mathbf{X}}_{e,2}}{{\mathbf{\Theta }}_e}{{\mathbf{Y}}_{e,2}}} \right) + {\text{Tr}}\left( {{{\mathbf{\Theta }}_e}{{\mathbf{Z}}_{e,2}}} \right)
   + {\text{Tr}}\left( {{\mathbf{\Theta }}_e^H{\mathbf{Z}}_{e,2}^H} \right) + {d_{e,2}} \leq {P_{th}},  \tag{\ref{P7_OF}{a}}  \label{P7_sinr}  \\
& \ \  a_{e,l}^2 + b_{e,l}^2 \leq 1, \tag{\ref{P7_OF}{b}}  \label{P7_coeff} \\
& \ \  0 \leq{a_{e,l}},{b_{e,l}} \leq 1, \tag{\ref{P7_OF}{c}}  \label{P7_01}  \\
& \ \  0 \leq {\alpha _{e,l}},{\beta _{e,l}} \leq 2\pi , \ \  \forall l, \tag{\ref{P7_OF}{d}}  \label{P7_phase}
\end{align}\end{small}where
\begin{small}\begin{align*}
{{\mathbf{X}}_{e,1}}\! \!=\! {{\mathbf{h}}_{id}}{\mathbf{h}}_{id}^H, \ \ {{\mathbf{X}}_{e,2}}\! \!=\! {{\mathbf{H}}_{ir}}{\mathbf{H}}_{ir}^H, \ \ {d_{e,2}}\! =\! {\text{Tr}}\left( {{\mathbf{H}}_{tr}^H{\mathbf{w}}{{\mathbf{w}}^H}{{\mathbf{H}}_{tr}}} \!\right),
\end{align*}\end{small}
\vspace{-1.7em}
\begin{small}
\begin{equation*}
{{\mathbf{Y}}_{e,1}} \! =\! {{\mathbf{Y}}_{e,2}} \! =\!  {{\mathbf{H}}_{ti}}{\mathbf{w}}{{\mathbf{w}}^H}{\mathbf{H}}_{ti}^H, \qquad {{\mathbf{Z}}_{e,2}} = {{\mathbf{H}}_{ti}}{\mathbf{w}}{{\mathbf{w}}^H}{{\mathbf{H}}_{tr}}{\mathbf{H}}_{ir}^H.
\vspace{-0.5em}
\end{equation*}\end{small}However, this problem is still not convex due to the objective function \eqref{P7_OF} and constraints \eqref{P7_coeff}-\eqref{P7_phase}.  By introducing new variables
${\boldsymbol{\alpha }_e} = \left[ {{a_{e,1}}{e^{j{\alpha _{e,1}}}},{a_{e,2}}{e^{j{\alpha _{e,2}}}}, \ldots ,{a_{e,L}}{e^{j{\alpha _{e,L}}}}} \right] \in {\mathbb{C}^{L \times 1}}$ and
${\boldsymbol{\beta }_e} = \left[ {{b_{e,1}}{e^{j{\beta _{e,1}}}},{b_{e,2}}{e^{j{\beta _{e,2}}}}, \ldots ,{b_{e,L}}{e^{j{\beta _{e,L}}}}} \right] \in {\mathbb{C}^{L \times 1}}$, we can reformulate the constraint \eqref{P7_coeff}-\eqref{P7_phase} as follows:
\begin{equation}\small
{\text{diag}}\left\{ {{\boldsymbol{\alpha }_e}{{\boldsymbol{\alpha }}_e^H} + {\boldsymbol{\beta }_e}{{\boldsymbol{\beta }}_e^H}} \right\} \leq {{\mathbf{1}}_L},
\end{equation}
which is now convex with respect to $\boldsymbol{\alpha }_e$ and $\boldsymbol{\beta}_e$. In addition, according to the matrix properties in \cite[Eq. (1.10.6)]{Matrix_Analysis}, we have
${\text{Tr}}\left( {{{\boldsymbol{\Phi }}_e^H}{{\mathbf{X}}_{e,1}}{\boldsymbol{\Phi }_e}{{\mathbf{Y}}_{e,1}}} \right) = {{\boldsymbol{\beta }}_e^H}{\boldsymbol{\Xi} _{e,1}}{\boldsymbol{\beta }_e}$ and
\begin{equation}\small
  {\text{Tr}}\left( {{{\mathbf{\Theta }}_e^H}{{\mathbf{X}}_{e,2}}{\mathbf{\Theta }_e}{{\mathbf{Y}}_{e,2}}} \right)\! + \!{\text{Tr}}\left( {{\mathbf{\Theta }_e}{{\mathbf{Z}}_{e,2}}} \right) \!+\! {\text{Tr}}\left( {{{\mathbf{\Theta }}_e^H}{\mathbf{Z}}_{e,2}^H} \right) \!+\! {d_{e,2}} = {{\boldsymbol{\alpha }}_e^H}{\boldsymbol{\Xi} _{e,2}}{\boldsymbol{\alpha }_e} \!+\! {\mathbf{z}}_{e,2}^H{{\boldsymbol{\alpha }}_e^ * } \!+\! {{\boldsymbol{\alpha }}_e^T}{{\mathbf{z}}_{e,2}}\! +\! {d_{e,2}},
\end{equation}
where ${\boldsymbol{\Xi} _{e,1}} = {{\mathbf{X}}_{e,1}} \odot {{\mathbf{Y}}_{e,1}}$, ${\boldsymbol{\Xi} _{e,2}} = {{\mathbf{X}}_{e,2}} \odot {{\mathbf{Y}}_{e,2}}$ and ${{\mathbf{z}}_{e,2}} = {\text{diag}}\left\{ {{{\mathbf{Z}}_{e,2}}} \right\}$. Therefore, the reflecting and refracting phase shifts optimization problem for maximizing the data rate can be reformulated as follows:

\vspace{-1.7em}
\begin{small}\begin{align}
 \mathop {\max }\limits_{\boldsymbol{\alpha}_e,\boldsymbol{\beta}_e}&\ \ {{\boldsymbol{\beta }}_e^H}{\boldsymbol{\Xi }_{e,1}}{\boldsymbol{\beta }_e} \label{P8_OF}\\
{\text{s.t.}} & \ \  {\text{diag}}\left\{ {{\boldsymbol{\alpha }_e}{{\boldsymbol{\alpha }}_e^H} + {\boldsymbol{\beta }_e}{{\boldsymbol{\beta }}_e^H}} \right\} \leq {{\mathbf{1}}_L},  \tag{\ref{P8_OF}{a}}  \label{P8_modulus} \\
& \ \  {{\boldsymbol{\alpha }}_e^H}{\boldsymbol{\Xi} _{e,2}}{\boldsymbol{\alpha }_e} + {\mathbf{z}}_{e,2}^H{{\boldsymbol{\alpha }}_e^ * } + {{\boldsymbol{\alpha }}_e^T}{{\mathbf{z}}_{e,2}} + {d_{e,2}} \!\leq\! {P_{th}}, \tag{\ref{P8_OF}{b}}  \label{P8_sinr}
\end{align}\end{small}so far, \eqref{P8_OF} is the only obstacle for this optimization, which can be approximated via the first-order Taylor expansion as follows:
\begin{equation}\small
{{\boldsymbol{\beta }}_e^H}{\boldsymbol{\Xi} _{e,1}}{\boldsymbol{\beta }_s} \geq 2\operatorname{Re} \left\{ {{{\boldsymbol{\beta }}_e^H}{\boldsymbol{\Xi} _{e,1}}{\boldsymbol{\tilde \beta }_e}} \right\} - {{{\boldsymbol{\tilde \beta }}}_e^H}{\boldsymbol{\Xi} _{e,1}}{\boldsymbol{\tilde \beta }_e},
\end{equation}
where ${{\boldsymbol{\tilde \beta }_e}}$ is the refracting phase shifts derived at the last iteration, therefore, the reflecting and refracting phase shifts optimization can be rewritten as follows:

\vspace{-1.7em}
\begin{small}\begin{align}
 \mathop {\max }\limits_{\boldsymbol{\alpha}_e,\boldsymbol{\beta}_e}&\ \ \operatorname{Re} \left\{ {{{\boldsymbol{\beta }}_e^H}{\boldsymbol{\Xi} _{e,1}}{\boldsymbol{\tilde \beta }_e}} \right\} \label{P9_OF}\\
{\text{s.t.}} & \ \  {\text{diag}}\left\{ {{\boldsymbol{\alpha }_e}{{\boldsymbol{\alpha }}_e^H} + {\boldsymbol{\beta }_e}{{\boldsymbol{\beta }}_e^H}} \right\} \leq {{\mathbf{1}}_L},  \tag{\ref{P9_OF}{a}}  \label{P9_modulus} \\
& \ \  {{\boldsymbol{\alpha }}_e^H}{\boldsymbol{\Xi} _{e,2}}{\boldsymbol{\alpha }_e} + {\mathbf{z}}_{e,2}^H{{\boldsymbol{\alpha }}^ * } + {{\boldsymbol{\alpha }}_e^T}{{\mathbf{z}}_{e,2}} + {d_{e,2}} \!\leq\! {P_{th}}, \tag{\ref{P9_OF}{b}}  \label{P9_sinr}
\end{align}\end{small}which now is a convex problem with convex constraints, and can be solved by CVX using QCQP. Therefore, the joint amplitudes and phase shifts of the ES-IOS are solved. In addition, the overall algorithm for maximizing the data rate for the ES-IOS is summarized in Algorithm 1.
\begin{algorithm}[t]
\caption{Gaussian randomization procedure}
\label{randomization}
\algsetup{linenosize=\footnotesize}
\footnotesize
\begin{algorithmic}
\STATE 1. Set a number of randomizations $G$,  given the SDR solution $\mathbf{X}$. \\
\STATE 2. Generate  $\boldsymbol{\xi}_g \sim \mathcal{N}\left(\mathbf{0}, \mathbf{X} \right)$, $g=1,2,...,G$ and construct a feasible point $\tilde{\boldsymbol{x}}_{g}=\operatorname{sgn}\left( {\boldsymbol{\xi}}_g \right)$.\\
\STATE 3. Determine $g^{\star}\! =\!  \arg \max\limits_{g=1, \ldots, G} \!  {\text{Tr}}\left( {{{{\mathbf{\Xi '}}}_{m,1}}\tilde{\boldsymbol{x}}_{g}{\tilde{\boldsymbol{x}}_{g}^T}} \right) $ and $[\boldsymbol{x}_{g^{\star}}]_{L+1}\!  =\! 1$ for the data rate maximization problem,  $g^{\star} = \arg \min\limits_{g=1, \ldots, G}  {\text{Tr}}\left( {{{{\mathbf{\Xi '}}}_{m,2}}\tilde{\boldsymbol{x}}_{g}{\tilde{\boldsymbol{x}}_{g}^T}} \right) $ and $[\boldsymbol{x}_{g^{\star}}]_{L+1}\!  =\! 1$ for the SI power minimization problem.
\STATE 4. Set $\mathbf{b}=[ \boldsymbol{x}_{g^{\star}}]_{1:L}$, calculate $\mathbf{a}_m=\frac{\mathbf{b}+\mathbf{1}_L}{2}$, and output $\mathbf{A}_m=\text{diag}\left\{ {\mathbf{a}_m} \right\}$.
\end{algorithmic}
\end{algorithm}
\\
\\
b) \textit{Optimizing reflecting and refracting phase shifts $\mathbf{\Theta '}_m$ and $\mathbf{\Phi '}_m$ for MS-IOS}

In the MS-IOS case, with a given beamforming vector $\mathbf{w}$, we reformulate the rate maximization problem as follows:

\vspace{-1.7em}
\begin{small}\begin{align}
 \mathop {\max }\limits_{\mathbf{\Theta}_m,\mathbf{\Phi}_m}&\ \ {\mathbf{h}}_{id}^H{\mathbf{\Phi }_m}{{\mathbf{H}}_{ti}}{\mathbf{w}}{{\mathbf{w}}^H}{\mathbf{H}}_{ti}^H{{\mathbf{\Phi }}_m^H}{{\mathbf{h}}_{id}} \label{P10_OF}\\
{\text{s.t.}} & \ \
  {\text{Tr}}\left( {\left( {{\mathbf{H}}_{tr}^H \!\! +\!\! {\mathbf{H}}_{ir}^H{\mathbf{\Theta }_m}{{\mathbf{H}}_{ti}}} \right){\mathbf{w}}{{\mathbf{w}}^H}\left( {{\mathbf{H}}_{tr}^H \!\!+\!\! {\mathbf{H}}_{ti}^H{{\mathbf{\Theta }}_m^H}{{\mathbf{H}}_{ir}}} \right)} \right)
   \leq {P_{th}},
\tag{\ref{P10_OF}{a}}  \label{P10_sinr}  \\
& \ \  {a_{m,l}} + {b_{m,l}} = 1, \tag{\ref{P10_OF}{b}}  \label{10_sum1} \\
& \ \  {a_{m,l}},{b_{m,l}} \in \left\{ {0,1} \right\}, \tag{\ref{P10_OF}{c}}  \label{10_01}  \\
& \ \ 0 \leq {\alpha _{m,l}},{\beta _{m,l}} \leq 2\pi,  \ \  \forall l, \tag{\ref{P10_OF}{d}}  \label{P10_phase}
\end{align}\end{small}however, this problem is non-convex since the binary variables are contained in the coefficient matrices. To tackle this problem, we further divide this problem into two sub-problems.

Firstly, we reconstruct the phase shifts coefficient matrices as ${{\mathbf{\Theta }}_m}\! \!=\! {{\mathbf{A}}_m}{{\mathbf{\Theta}_m'}}$ and ${{\mathbf{\Phi }}_m} \!=\! {{\mathbf{B}}_m}{\mathbf{\Phi}_m'}$, where ${{\mathbf{A}}_m} = {\text{diag}}\left\{ {{a_{m,1}},{a_{m,2}}, \ldots ,{a_{m,L}}} \right\} \in {\mathbb{C}^{L \times L}}$ and ${{\mathbf{B}}_m} = {\text{diag}}\left\{ {{b_{m,1}},{b_{m,2}}, \ldots ,{b_{m,L}}} \right\} \in {\mathbb{C}^{L \times L}}$ indicate the elements for reflecting and refracting. ${\mathbf{\Theta}_m'} = {\text{diag}}\left\{ {{e^{j{\alpha _{m,1}}}},{e^{j{\alpha _{m,2}}}}, \ldots ,{e^{j{\alpha _{m,L}}}}} \right\} \in {\mathbb{C}^{L \times L}}$ and ${\mathbf{\Phi}_m'} = {\text{diag}}\left\{ {{e^{j{\beta _{m,1}}}},{e^{j{\beta _{m,2}}}}, \ldots ,{e^{j{\beta _{m,L}}}}} \right\} \in {\mathbb{C}^{L \times L}}$ are the reflecting and refracting phase shifts matrices, respectively. Thereby, the channels can be reformulated as follows:
\begin{equation}\small
{{\mathbf{h}}_{d,m}} = {\mathbf{h}}_{id}^H\left( {{\mathbf{I}} - {{\mathbf{A}}_m}} \right){\mathbf{\Phi}_m'}{{\mathbf{H}}_{ti}},
\end{equation}
\begin{equation}\small
{{\mathbf{H}}_{r,m}} = {\mathbf{H}}_{tr}^H + {\mathbf{H}}_{ir}^H{{\mathbf{A}}_m}{\mathbf{\Phi}_m'}{{\mathbf{H}}_{ti}},
\end{equation}
then the problem with fixed mode selection matrix $\mathbf{A}_m$ and $\mathbf{B}_m$ can be reformulated as follows:

\vspace{-1.7em}
\begin{small}\begin{align}
 \mathop {\max }\limits_{\mathbf{\Theta '}_m,\mathbf{\Phi '}_m}&\  {\mathbf{h}}_{id}^H\left( {{\mathbf{I}} - {{\mathbf{A}}_m}} \right){\mathbf{\Phi}_m'}{{\mathbf{H}}_{ti}}{\mathbf{w}}{{\mathbf{w}}^H}{\mathbf{H}}_{ti}^H{\mathbf{\Phi}_m'^H}\left( {{\mathbf{I}} - {{\mathbf{A}}_m}}\right)^H {{\mathbf{h}}_{id}} \label{P11_OF}\\
{\text{s.t.}} \ & \
  {\text{Tr}}\left(\! {\left(\! {{\mathbf{H}}_{tr}^H \!\!+\!\! {\mathbf{H}}_{ir}^H\!{{\mathbf{A}}_m}\!{\mathbf{\Theta}_m'}\!{{\mathbf{H}}_{ti}}} \right){\mathbf{w}}\!{{\mathbf{w}}^H}\!\left( {{{\mathbf{H}}_{tr}}\!\! +\!\! {\mathbf{H}}_{ti}^H\!{\mathbf{\Theta}_m'^H}\!{\mathbf{A}}_m^H\!{{\mathbf{H}}_{ir}}}\! \right)} \!\right)   \leq {P_{th}},
  \tag{\ref{P11_OF}{a}}  \label{P11_sinr}  \\
& \  0 \leq {\alpha _{m,l}},{\beta _{m,l}} \leq 2\pi,  \ \  \forall l. \tag{\ref{P11_OF}{b}}  \label{P11_phase}
\end{align}\end{small}Note that by introducing ${\boldsymbol{\alpha }_m} \!= \! \left[ {{e^{j{\alpha _{m,1}}}},{e^{j{\alpha _{m,2}}}}, \ldots ,{e^{j{\alpha _{m,L}}}}} \right] \in {\mathbb{C}^{L \times 1}}$ and
${{\boldsymbol{\beta }}_m} = \left[ {{e^{j{\beta _{m,1}}}},{e^{j{\beta _{m,2}}}}, \ldots } \right.,$ $\left. {{e^{j{\beta _{m,L}}}}} \right] \in {\mathbb{C}^{L \times 1}}$, the constraint \eqref{P11_phase} can be reformulated as
\begin{equation}\small
{\text{diag}}\left\{ {\boldsymbol{\alpha }_m}{{\boldsymbol{\alpha }}_m^H}  \right\} = {{\mathbf{1}}_L},
\end{equation}
\begin{equation}\small
{\text{diag}}\left\{ {\boldsymbol{\beta }_m}{{\boldsymbol{\beta }}_m^H}  \right\} = {{\mathbf{1}}_L},
\end{equation}
and the problem can be reformulated as

\vspace{-1.7em}
\begin{small}\begin{align}
 \mathop {\max }\limits_{\boldsymbol{\alpha}_m,\boldsymbol{\beta}_m}&\ \ {{\boldsymbol{\beta }}_m^H}{\boldsymbol{\Xi}_{m,1}}{\boldsymbol{\beta }_m} \label{P12_OF}\\
{\text{s.t.}} & \ \ {{\boldsymbol{\alpha }}_m^H}{\boldsymbol{\Xi} _{m,2}}{\boldsymbol{\alpha }_m} \! +\!  {\mathbf{z}}_{m,2}^H{{\boldsymbol{\alpha }}_m^ * }\!  +\!  {{\boldsymbol{\alpha }}_m^T}{{\mathbf{z}}_{m,2}} + {d_{m,2}} \!\leq\! {P_{th}}, \tag{\ref{P12_OF}{a}}  \label{P12_sinr}\\
& \ \  {\text{diag}}\left\{ {\boldsymbol{\alpha }_m}{{\boldsymbol{\alpha }}_m^H}  \right\} = {{\mathbf{1}}_L}, \tag{\ref{P12_OF}{b}} \label{P12_modulus1}\\
& \ \  {\text{diag}}\left\{ {\boldsymbol{\beta }_m}{{\boldsymbol{\beta }}_m^H}  \right\} = {{\mathbf{1}}_L}, \tag{\ref{P12_OF}{c}} \label{P12_modulus2}
\end{align}\end{small}
where
\begin{small}\begin{equation*}
{{\mathbf{\Xi }}_{m,1}} = {{\mathbf{X}}_{m,1}} \odot {{\mathbf{Y}}_{m,1}}, \qquad\qquad {{\mathbf{\Xi }}_{m,2}} = {{\mathbf{X}}_{m,2}} \odot {{\mathbf{Y}}_{m,2}},\vspace{-1.5em}
\end{equation*}\end{small}
\begin{small}\begin{equation*}
{{\mathbf{X}}_{m,1}} \!\!=\! {\left( {{\mathbf{I}}\! \!-\!\! {{\mathbf{A}}_m}} \right)^H}{{\mathbf{h}}_{id}}{\mathbf{h}}_{id}^H\left( {{\mathbf{I}} \!\!-\!\! {{\mathbf{A}}_m}} \right),\  {{\mathbf{X}}_{m,2}}\! \!=\! {\mathbf{A}}_o^H{{\mathbf{H}}_{ir}}{\mathbf{H}}_{ir}^H{{\mathbf{A}}_m},\vspace{-1.5em}
\end{equation*}\end{small}
\begin{small}\begin{equation*}
{{\mathbf{Y}}_{m,1}} = {{\mathbf{H}}_{ti}}{\mathbf{H}}_{ti}^H, \qquad\qquad\qquad\ {{\mathbf{Y}}_{m,2}} = {{\mathbf{H}}_{ti}}{\mathbf{w}}{{\mathbf{w}}^H}{\mathbf{H}}_{ti}^H,\vspace{-1.5em}
\end{equation*}\end{small}
\begin{small}\begin{equation*}
{{\mathbf{Z}}_{m,2}}\! \!=\! {{\mathbf{H}}_{ti}}{\mathbf{w}}{{\mathbf{w}}^H}{{\mathbf{H}}_{tr}}{\mathbf{H}}_{ir}^H{{\mathbf{A}}_m},\quad  {d_{m,2}}\! =\! {\text{Tr}}\left( {{\mathbf{H}}_{tr}^H{\mathbf{w}}{{\mathbf{w}}^H}{{\mathbf{H}}_{tr}}} \!\right),
\end{equation*}\end{small}and $\mathbf{z}_{m,2}=\text{diag}\{{\mathbf{Z}_{m,2}} \}$,
however, this problem is non-convex due to \eqref{P12_OF}, \eqref{P12_modulus1} and \eqref{P12_modulus2}. By applying the first-order Taylor expansion to \eqref{P12_OF}, and relaxation to \eqref{P12_modulus1} and \eqref{P12_modulus2}, the problem can be transformed as follows:

\vspace{-1.7em}
\begin{small}\begin{align}
 \mathop {\max }\limits_{\boldsymbol{\alpha}_m,\boldsymbol{\beta}_m}&\ \ \operatorname{Re} \left\{ {{\boldsymbol{\beta }}_m^H}{\boldsymbol{\Xi}_{m,1}}{\boldsymbol{\tilde \beta }_m} \label{P13_OF}  \right\} \\
{\text{s.t.}} & \ \ {{\boldsymbol{\alpha }}_m^H}{\boldsymbol{\Xi} _{m,2}}{\boldsymbol{\alpha }_m} + {\mathbf{z}}_{m,2}^H{{\boldsymbol{\alpha }}_m^ * } + {{\boldsymbol{\alpha }}_m^T}{{\mathbf{z}}_{m,2}} + {d_{m,2}} \!\leq\! {P_{th}}, \tag{\ref{P13_OF}{a}}  \label{P13_sinr}\\
& \ \  {\text{diag}}\left\{ {\boldsymbol{\alpha }_m}{{\boldsymbol{\alpha }}_m^H}  \right\} \leq {{\mathbf{1}}_L}, \tag{\ref{P13_OF}{b}} \label{P13_modulus1}\\
& \ \  {\text{diag}}\left\{ {\boldsymbol{\beta }_m}{{\boldsymbol{\beta }}_m^H}  \right\} \leq {{\mathbf{1}}_L}, \tag{\ref{P13_OF}{c}} \label{P13_modulus2}
\end{align}\end{small}where $\boldsymbol{\tilde \beta }_o$ is the value at the last iteration, now this problem can be solved. In the following subsection, the mode selection for the case with MS-IOS will be discussed.

\subsection{Optimizing mode selection $\mathbf{A}_m$ with given $\mathbf{w}$, $\mathbf{\Theta '}_m$ and $\mathbf{\Phi '}_m$ for MS-IOS}
In this subsection, we aim to optimize $\mathbf{A}_m$ with given phase shifts and beamforming vector, for the case with MS-IOS. The problem is formulated as follows:

\vspace{-1em}
\begin{small}\begin{align}
 \mathop {\max }\limits_{\mathbf{A}_m}&\ \ {\mathbf{h}}_{id}^H\left( {{\mathbf{I}}\! - \!{{\mathbf{A}}_m}} \right){\mathbf{\Phi}_m'}\!{{\mathbf{H}}_{ti}}{\mathbf{w}}{{\mathbf{w}}^H}{\mathbf{H}}_{ti}^H{\mathbf{\Phi}_m'^H}\left( {{\mathbf{I}} \!-\! {{\mathbf{A}}_m}}\right)^H {{\mathbf{h}}_{id}} \label{P14_OF}\\
{\text{s.t.}} & \ \
  {\text{Tr}}\left( {\left( {{\mathbf{H}}_{tr}^H \!\!+\!\! {\mathbf{H}}_{ir}^H\!{{\mathbf{A}}_m}\!{\mathbf{\Theta}_m'}\!{{\mathbf{H}}_{ti}}} \right){\mathbf{w}}\!{{\mathbf{w}}^H}\!\left( {{{\mathbf{H}}_{tr}}\!\! +\!\! {\mathbf{H}}_{ti}^H\!{\mathbf{\Theta}_m'^H}\!{\mathbf{A}}_m^H\!{{\mathbf{H}}_{ir}}}\! \right)} \right)\leq {P_{th}},  \tag{\ref{P14_OF}{a}}  \label{P14_sinr}  \\
& \ \ {a_{m,l}} + {b_{m,l}} = 1, \tag{\ref{P14_OF}{b}} \label{P14_sum1}\\
& \ \ {a_{m,l}},{b_{m,l}} \in \left\{ {0,1} \right\}, \tag{\ref{P14_OF}{c}} \label{P14_01}
\end{align}\end{small}however, this problem is difficult to tackle due to the binary variables. Firstly, we need to reconstruct the problem, by introducing a vector  ${{\mathbf{a}}_m} = {\left[ {{a_{m,1}},{a_{m,2}}, \ldots ,{a_{m,L}}} \right]} \in {\mathbb{C}^{L \times 1}}$, we can rewrite the problem as follows:

\vspace{-1em}
\begin{small}\begin{align}
 \mathop {\max }\limits_{\mathbf{a}_m}&\ \ {\mathbf{a}}_m^T{{\mathbf{\Xi }}_{m,1}}{{\mathbf{a}}_m} - 2\operatorname{Re} \left\{ {{\mathbf{a}}_m^T{{\mathbf{w}}_{m,1}}} \right\} + {d_{m,1}} \label{P15_OF}   \\
{\text{s.t.}} & \ \ {\mathbf{a}}_m^T{{\mathbf{\Xi }}_{m,2}}{{\mathbf{a}}_m} + 2\operatorname{Re} \left\{ {{\mathbf{a}}_m^T{{\mathbf{w}}_{m,2}}} \right\} + {d_{m,2}} \leq {P_{th}}, \tag{\ref{P15_OF}{a}}  \label{P15_sinr}\\
& \ \ {a_{m,l}} \in \left\{ {0,1} \right\}, \tag{\ref{P15_OF}{c}} \label{P15_01}
\end{align}\end{small}
where
\begin{small}\begin{equation*}
{{\mathbf{\Xi }}_{m,1}} = {{\mathbf{U}}_{m,1}} \odot {{\mathbf{V}}_{m,1}}, \qquad\qquad {{\mathbf{\Xi }}_{m,2}} = {{\mathbf{U}}_{m,2}} \odot {{\mathbf{V}}_{m,2}},\vspace{-1.5em}
\end{equation*}\end{small}
\begin{small}\begin{equation*}
{{\mathbf{U}}_{m,1}} = {{\mathbf{h}}_{id}}{\mathbf{h}}_{id}^H,\qquad\qquad {{\mathbf{V}}_{m,1}} = {{{\mathbf{\Phi '}_m}}}{{\mathbf{H}}_{ti}}{\mathbf{w}}{{\mathbf{w}}^H}{\mathbf{H}}_{ti}^H{{{\mathbf{\Phi '}}}_m^H},\vspace{-1.5em}
\end{equation*}\end{small}
\begin{small}\begin{equation*}
{{\mathbf{U}}_{m,2}} = {{\mathbf{H}}_{ir}}{\mathbf{H}}_{ir}^H,\qquad\qquad{{\mathbf{V}}_{m,2}} = {{{\mathbf{\Theta '}}}_m}{{\mathbf{H}}_{ti}}{\mathbf{w}}{{\mathbf{w}}^H}{\mathbf{H}}_{ti}^H{{{\mathbf{\Theta '}}}_m^H},\vspace{-1.5em}
\end{equation*}\end{small}
\begin{small}\begin{equation*}
{{\mathbf{W}}_{m,1}} = {\text{diag}}\left\{ {{\mathbf{h}}_{id}^H} \right\}{{{\mathbf{\Phi '}}}_m}{{\mathbf{H}}_{ti}}{\mathbf{w}}{{\mathbf{w}}^H}{\mathbf{H}}_{ti}^H{{{\mathbf{\Phi '}}}_m}^H{{\mathbf{h}}_{id}},\vspace{-1.5em}
\end{equation*}\end{small}
\begin{small}\begin{equation*}
{{\mathbf{W}}_{m,2}} = {{{\mathbf{\Theta '}}}_m}{{\mathbf{H}}_{ti}}{\mathbf{w}}{{\mathbf{w}}^H}{{\mathbf{H}}_{tr}}{\mathbf{H}}_{ir}^H,\vspace{-1.5em}
\end{equation*}\end{small}
\begin{small}\begin{equation*}
{d_{m,1}}\! \!=\!\! {\mathbf{h}}_{id}^H{{{\mathbf{\Phi '}}}_m}{{\mathbf{H}}_{ti}}{\mathbf{w}}{{\mathbf{w}}^H}{\mathbf{H}}_{ti}^H{{{\mathbf{\Phi '}}}_m^H}{{\mathbf{h}}_{id}},\ {d_{m,2}} \!\!=\! {\text{Tr}}\left( {{\mathbf{H}}_{tr}^H{\mathbf{w}}{{\mathbf{w}}^H}{{\mathbf{H}}_{tr}}}\! \right).
\end{equation*}\end{small}
However, \eqref{P15_OF} is still not convex due to the binary variables. By reconstructing the expressions ${\mathbf{a}}_m^T{{\mathbf{\Xi }}_{m,1}}{{\mathbf{a}}_m} = \sum\limits_{{l_1} = 1}^L {\sum\limits_{{l_2} = 1}^L {{{\left[ {{{\mathbf{\Xi }}_{m,1}}} \right]}_{{l_1},{l_2}}}{a_{{l_1}}}} } {a_{{l_2}}}$, and $\operatorname{Re} \left\{ {{\mathbf{a}}_m^T{{\mathbf{w}}_{m,1}}} \right\}\! =\! \sum\limits_{{l_1}\! = \!1}^L {{a_{{l_1}}}} {\left[ {\operatorname{Re} \left\{ {{{\mathbf{w}}_{m,1}}} \right\}} \right]_{{l_1}}}$. In addition, by introducing slack variables ${{\mathbf{S}}_m} = \left\{ {\left. {{s_{{l_1},{l_2}}}} \right|{l_1},{l_2} = 1,2, \ldots ,L} \right\} \in {\mathbb{C}^{L \times L}}$ we can reformulate the problem as follows:

\vspace{-1.7em}
\begin{small}\begin{align}
 \mathop {\max }\limits_{\mathbf{a}_m,\mathbf{S}_m}&\ \ \sum\limits_{{l_1} = 1}^L {\sum\limits_{{l_2} = 1}^L {{{\left[ {{{\mathbf{\Xi }}_{m,1}}} \right]}_{{l_1},{l_2}}}} } {s_{{l_1},{l_2}}} - 2{a_{{l_1}}}\sum\limits_{{l_1} = 1}^L {{{\left[ {\operatorname{Re} \left\{ {{{\mathbf{w}}_{m,1}}} \right\}} \right]}_{{l_1}}}}  \label{P16_OF}   \\
{\text{s.t.}} & \ \
  \sum\limits_{{l_1} = 1}^L {\sum\limits_{{l_2} = 1}^L {{{\left[ {{{\mathbf{\Xi }}_{m,2}}} \right]}_{{l_1},{l_2}}}} } {s_{{l_1},{l_2}}} + 2{a_{{l_1}}}\sum\limits_{{l_1} = 1}^L  {\left[ {\operatorname{Re} \left\{ {{{\mathbf{w}}_{m,2}}} \right\}} \right]_{{l_1}}}
  \ + {d_{m,2}} \leq {P_{th}},  \tag{\ref{P16_OF}{a}}  \label{P16_sinr}\\
& \ \ {a_{m,l}} \in \left\{ {0,1} \right\}, \tag{\ref{P16_OF}{b}} \label{P16_01}\\
& \ \ {s_{{l_1},{l_2}}} \leq {a_{{l_1}}}, \tag{\ref{P16_OF}{c}} \label{P16_c1}\\
& \ \ {s_{{l_1},{l_2}}} \leq {a_{{l_2}}},  \tag{\ref{P16_OF}{d}} \label{P16_c2}\\
& \ \ {s_{{l_1},{l_2}}} \geq {a_{{l_1}}} + {a_{{l_2}}} - 1 , \tag{\ref{P16_OF}{e}} \label{P16_c3}
\end{align}\end{small}this problem can be solved by CVX, however, is NP-hard \cite{BQP_np} thus this problem is intractable for a large $L$.

\begin{algorithm}[t]
\caption{Data rate maximizations for both ES-IOS and MS-IOS}
\label{Maximizing}
\algsetup{linenosize=\footnotesize}
\footnotesize
\begin{algorithmic}
\STATE {Initialize the beamforming vector ${\mathbf{w}}^{0}$, the reflecting, refracting amplitudes and phase shifts ${\mathbf{\Theta}_e ^{0}}$, ${\mathbf{\Phi}_e ^{0}}$ of the ES-IOS,  the phase shifts matrix of MS-IOS  ${\mathbf{\Theta}_m ^{0}}$, ${\mathbf{\Phi}_m ^{0}}$, as well as the mode selection matrix of the MS-IOS $\mathbf{A}_m^0$,
compute $R_{d,e}\left( {{\mathbf{w}}^0,\!{{\mathbf{\Theta }}_e^0},\!{{\mathbf{\Phi }}_e^0}} \right)$ and $R_{d,m}\left( {{\mathbf{w}}^0,\!{{\mathbf{\Theta }}_m^0},\!{{\mathbf{\Phi }}_m^0},{\mathbf{A}_m^0}} \right)$, respectively. Set $t=0$ and the accuracy for iteration $\varepsilon_1$.}
\REPEAT
\STATE  1. Given ${\mathbf{w}}^{ t }$, $\mathbf{A}_m^0$, ${\mathbf{\Theta}_e^{t}}$,  ${\mathbf{\Phi}_e^{t}}$, ${\mathbf{\Theta}_m^{t}}$ and ${\mathbf{\Phi}_m^{t}}$, compute the data rates  as $R_{d,e}\left( {{\mathbf{w}}^{t}, {\mathbf{\Theta}_e ^{ t}}, {\mathbf{\Phi}_e ^{ t}}} \right)$ and $R_{d,m}\left( {{\mathbf{w}}^{t}, {\mathbf{\Theta}_m ^{ t}}, {\mathbf{\Phi}_m^{ t}}},{\mathbf{A}_m^t} \right)$, respectively.\\
2. Given ${\mathbf{\Theta}_e^{t}}$, ${\mathbf{\Phi}_e ^{t}}$ for the case with ES-IOS and given $\mathbf{A}_m^t$, ${\mathbf{\Theta}_m^{t}}$ and ${\mathbf{\Phi}_m ^{t}}$ for the case with MS-IOS, optimize ${\mathbf{w}}^{t+1}$  by solving problem \eqref{P5_OF}.\\
3. Given ${\mathbf{w}}^{t+1 }$, optimize ${\mathbf{\Theta} _e^{ t+1 }}$ and ${\mathbf{\Phi} _e^{ t+1 }}$ by solving problem \eqref{P9_OF} for the ES-IOS. As for the MS-IOS, optimize ${\mathbf{\Theta} _m^{ t+1 }}$ and ${\mathbf{\Phi} _m^{ t+1 }}$ with given $\mathbf{w}^{t+1}$ and $\mathbf{A}_m^t$ for the case with MS-IOS, by solving problem \eqref{P13_OF}. \\
4. By solving \eqref{P161_OF} for the case with MS-IOS, optimize $\mathbf{X}$ given $\mathbf{w}^{t+1}$, $\mathbf{\Theta}_m^{t+1}$ and $\mathbf{\Phi}_m^{t+1}$.\\
5. For the case with MS-IOS, retrieve $\mathbf{A}_m^{t+1}$ from $\mathbf{X}$ using Gaussian randomization procedure. \\
6. Compute the data rates $R_{d,e}\left( {{\mathbf{w}}^{t+1}, {{\mathbf{\Theta }}_e^{t+1}},{{\mathbf{\Phi}}_e^{t+1}}} \right)$ and $R_{d,m}\left( {{\mathbf{w}}^{t+1}, {{\mathbf{\Theta }}_m^{t+1}},{{\mathbf{\Phi}}_m^{t+1}},\mathbf{A}_m^t} \right)$ for both the IOSs.\\
7. Set $t = t+1$.
\UNTIL {$\frac{{\left| {{R_{d,e}}\left( {{\mathbf{w}}^{t + 1},{\mathbf{\Theta }}_e^{t + 1},{\mathbf{\Phi }}_e^{t + 1}} \right) - {R_{d,e}}\left( {{\mathbf{w}}^t,{{\mathbf{\Theta }}_e^t},{{\mathbf{\Phi }}_e^t}} \right)} \right|}}{{{R_{d,e}}\left( {{\mathbf{w}}^{t + 1}, {{\mathbf{\Theta }}_e^{t + 1}},{{\mathbf{\Phi }}_e^{t + 1}}} \right)}} \leq \varepsilon_1$  or  $\frac{{\left| {{R_{d,m}}\left( {{\mathbf{w}}^{t + 1},{\mathbf{\Theta }}_m^{t + 1},{\mathbf{\Phi }}_m^{t + 1}, \mathbf{A}_m^{t+1}} \right) - {R_{d,m}}\left( {{\mathbf{w}}^t,{{\mathbf{\Theta }}_m^t},{{\mathbf{\Phi }}_m^t},\mathbf{A}_m^t} \right)} \right|}}{{{R_{d,m}}\left( {{\mathbf{w}}^{t + 1},{{\mathbf{\Theta }}_m^{t + 1}},{{\mathbf{\Phi }}_m^{t + 1}},\mathbf{A}_m^{t+1}} \right)}} \leq \varepsilon_1$.}
\end{algorithmic}
\end{algorithm}

Thus, we transform the above problem  using the SDR method. By introducing a new binary variable ${\mathbf{b}} = 2{\mathbf{a}_m} - {{\mathbf{1}}_L} \in {\mathbb{R}^{L \times 1}}$, where ${\mathbf{b}} = {\left[ {{b_1},{b_2}, \ldots ,{b_L}} \right]^T}$ and ${b_l} \in \left\{ { - 1,1} \right\}$, then the objective function of the rate maximization problem is given by
\begin{equation}\small
 \frac{1}{4}\left( {{\text{Tr}}\left( {{{\mathbf{\Xi }}_{m,1}}{\mathbf{B}}} \right) + 2\operatorname{Re} \left\{ {{{\mathbf{h}}^T}{\mathbf{b}}} \right\}} \right) + {c_{m,1}},
\end{equation}
where $\mathbf{B}=\mathbf{b}\mathbf{b}^T$. Thereby, by introducing another binary slack variable ${\mathbf{x}} = {\left[ {{\mathbf{b}};1} \right]}\in {\mathbb{R}^{\left( {L + 1} \right) \times 1}}$, we can reconstruct the objective function  and the SI power as
\begin{equation}\small
 \frac{1}{4}{\text{Tr}}\left( {{{{\mathbf{\Xi '}}}_{m,1}}{\mathbf{X}}} \right) + {c_{m,1}},
\end{equation}
\begin{equation}\small
 \frac{1}{4}{\text{Tr}}\left( {{{{\mathbf{\Xi '}}}_{m,2}}{\mathbf{X}}} \right) + {c_{m,2}},
\end{equation}
respectively.
The derivation is given in Appendix A.

Thus, the problem can be transformed as follows:

\vspace{-1.7em}
\begin{small}\begin{align}
 \mathop {\max }\limits_{\mathbf{X}}&\ \ \frac{1}{4}{\text{Tr}}\left( {{{{\mathbf{\Xi '}}}_{m,1}}{\mathbf{X}}} \right) + {c_{m,1}}  \label{P161_OF}   \\
{\text{s.t.}} & \ \ \frac{1}{4}{\text{Tr}}\left( {{{{\mathbf{\Xi '}}}_{m,2}}{\mathbf{X}}} \right) + {c_{m,2}} \leq\!  {P_{th}}, \tag{\ref{P161_OF}{a}}  \label{P161_sinr}\\
& \ \ {\text{diag}}\left\{ {\mathbf{X}} \right\} = {{\mathbf{1}}_{L + 1}}, \tag{\ref{P161_OF}{b}} \label{P161_1L}\\
& \ \  \text{Rank}\left( {\mathbf{X}} \right) = 1,  \tag{\ref{P161_OF}{c}} \label{P161_rank}
\end{align}\end{small}however, the rank-one constraint renders the problem infeasible, thus by dropping this constraint, we can optimize the problem by utilizing the SDR method. Then using the Gaussian randomization procedure, we can extract $\mathbf{a}_m$ from $\mathbf{X}$, the details of the Gaussian  randomization procedure is provided in Algorithm 1.

\subsection{Overall Algorithm and  complexity analysis}
In a nutshell, the overall alternative optimization algorithm for maximizing the data rate is provided in Algorithm 2.  The QCQP method is applied to solve the beamforming vector, the reflecting, refracting amplitudes and phase shifts optimization for the case with the ES-IOS and phase shifts optimization for the case with the MS-IOS.  Furthermore, the SDR method and Gaussian randomization procedure are exploited for solving the element mode selection for  MS-IOS. Every step in Algorithm 2 guarantees the objective function increase monotonically; therefore, the proposed algorithm's convergence is guaranteed.  It is worth pointing out that the complexity of the QCQP method for beamforming optimization is on the order of  $\mathcal{O}\left( {M^3} \right)$ \cite{convexopt},  and the computational complexities of the QCQP method for optimizing phase shifts matrices for  both the ES-IOS and MS-IOS are $\mathcal{O}\left({L^3} \right)$. In addition, the computational complexities of the SDR method and Gaussian randomization procedure for the mode selection optimization with the MS-IOS are $\mathcal{O}\left( {L^{3.5}} \right)$ \cite{sdrcomplexity} and $\mathcal{O}\left( {G} \right)$, respectively. Therefore, the overall complexity order of the proposed system amounts to
$\mathcal{O}\left( {{N_{ite}} \times \max \left\{ {M^3,{L^3}} \right\}} \right)$ and $\mathcal{O}\left( {{N_{ite}} \times \max \left\{ {M^3,{L^{3.5}},G} \right\}} \right)$, respectively, where $N_{ite}$ is the number of optimization iterations and $G$ is the number for randomizations.

\section{Minimizing the SI Power}
To minimize  the SI power, we notice that the problem is difficult to solve since the reflecting, refracting amplitudes/mode selection and  phase shifts $\mathbf{\Theta}_x$, $\mathbf{\Phi}_x$ and the beamforming vector $\mathbf{w}$ are coupled. Therefore, we propose an alternating algorithm to solve them iteratively.

\subsection{Optimizing beamforming vector $\mathbf{w}$ for given $\mathbf{\Theta}_x$ and $\mathbf{\Phi}_x$}

Firstly, with given reflecting, refracting amplitudes/mode selection and phase shifts $\mathbf{\Theta}_x$ and $\mathbf{\Phi}_x$ for ES-IOS and MS-IOS, the SI power minimization problem can be reformulated as follows:

\vspace{-1.7em}
\begin{small}\begin{align}
 \mathop {\min }\limits_{\mathbf{w}}&\ \ {\left\| {{{\mathbf{H}}_{r,x}}{\mathbf{w}}{{\mathbf{w}}^H}{\mathbf{H}}_{r,x}^H} \right\|_F} \label{P17_OF}\\
{\text{s.t.}}  & \ \   {\text{Tr}}\left( {{\mathbf{w}}{{\mathbf{w}}^H}} \right) \leq {P_{\max }},  \tag{\ref{P17_OF}{a}}  \label{P17_Power} \\
& \ \  {\log _2}\left( {1 + \frac{{{{\mathbf{h}}_{d,x}}{\mathbf{w}}{{\mathbf{w}}^H}{\mathbf{h}}_{d,x}^H}}{{\sigma _d^2}}} \right) \geq {R_{th}}, \tag{\ref{P17_OF}{b}}  \label{P17_sinr}
\end{align}\end{small}which can be rewritten as follows:

\vspace{-1.7em}
\begin{small}\begin{align}
 \mathop {\min }\limits_{\mathbf{w}}&\ \ {{\mathbf{w}}^H}{\mathbf{H}}_{r,x}^H{{\mathbf{H}}_{r,x}}{\mathbf{w}} \label{P18_OF}\\
{\text{s.t.}}  & \ \   {{\mathbf{w}}^H}{\mathbf{w}} \leq {P_{\max }},  \tag{\ref{P18_OF}{a}}  \label{P18_Power} \\
& \ \  {{\mathbf{w}}^H}{\mathbf{h}}_{d,x}^H{{\mathbf{h}}_{d,x}}{\mathbf{w}} \geq\left( {{2^{{R_{th}}}} - 1} \right)\sigma _d^2. \tag{\ref{P18_OF}{b}}  \label{P18_sinr}
\end{align}\end{small}We notice that constraint \eqref{P18_sinr} renders this problem non-convex, by using the first-order Taylor expansion, this problem can be reformulated as a convex optimization problem:

\vspace{-1.7em}
\begin{small}\begin{align}
 \mathop {\min }\limits_{\mathbf{w}}&\ \ {{\mathbf{w}}^H}{\mathbf{H}}_{r,x}^H{{\mathbf{H}}_{r,x}}{\mathbf{w}} \label{P19_OF}\\
{\text{s.t.}}  & \ \   {{\mathbf{w}}^H}{\mathbf{w}} \leq {P_{\max }},  \tag{\ref{P19_OF}{a}}  \label{P19_Power} \\
& \ \  2\operatorname{Re} \left( {{{\mathbf{w}}^H}{\mathbf{h}}_d^H{{\mathbf{h}}_d}{\mathbf{\tilde w}}} \right)\! \geq\! {{{\mathbf{\tilde w}}}^H}{\mathbf{h}}_{d,x}^H{{\mathbf{h}}_{d,x}}{\mathbf{\tilde w}} \!+\! \left( {{2^{{R_{th}}}}\! -\! 1} \right)\sigma _d^2, \tag{\ref{P19_OF}{b}}  \label{P19_sinr}
\end{align}\end{small}which is a QCQP problem  and can be solved.

\subsection{Optimizing reflecting and refracting amplitudes/mode selection and phase shifts $\mathbf{\Theta}_x$ and $\mathbf{\Phi}_x$ for given $\mathbf{w}$}
Firstly, we aim to optimize reflecting, refracting  amplitudes/ mode selection and phase shifts $\mathbf{\Theta}$ and $\mathbf{\Phi}$, for given beamforming vector $\mathbf{w}$.
\\ \\
a) \textit{Optimizing reflecting and refracting amplitudes and phase shifts $\mathbf{\Theta}_e$ and $\mathbf{\Phi}_e$ for ES-IOS}

For the ES-IOS, the SI power minimization problem can be formulated as follows:

\vspace{-1.7em}
\begin{small}\begin{align}
 \mathop {\min }\limits_{\mathbf{\Theta}_e,\mathbf{\Phi}_e}&\ \  {\text{Tr}}\left( {\left( {{\mathbf{H}}_{tr}^H \!\! +\!\! {\mathbf{H}}_{ir}^H{\mathbf{\Theta }_e}{{\mathbf{H}}_{ti}}} \right){\mathbf{w}}\!{{\mathbf{w}}^H}\!\left( {{\mathbf{H}}_{tr}^H \!\!+\!\! {\mathbf{H}}_{ti}^H{{\mathbf{\Theta }}_e^H}{{\mathbf{H}}_{ir}}} \right)} \right) \label{P20_OF}\\
{\text{s.t.}} & \ \  {\mathbf{h}}_{id}^H{\mathbf{\Phi }_e}{{\mathbf{H}}_{ti}}{\mathbf{w}}{{\mathbf{w}}^H}{\mathbf{H}}_{ti}^H{{\mathbf{\Phi }}_e^H}{{\mathbf{h}}_{id}} \geq\left( {{2^{{R_{th}}}} - 1} \right)\sigma _d^2, \tag{\ref{P20_OF}{a}}  \label{P20_sinr} \\
& \ \ a_{e,l}^2 + b_{e,l}^2 \leq 1, \tag{\ref{P20_OF}{b}}  \label{P20_1}\\
& \ \ 0 \leq{a_{e,l}},{b_{e,l}} \leq 1, \tag{\ref{P20_OF}{c}}  \label{P20_01}\\
& \ \ 0 \leq {\alpha _{e,l}},{\beta _{e,l}} \leq 2\pi , \ \  \forall l. \tag{\ref{P20_OF}{d}}  \label{P20_phase}
\end{align}\end{small}Following the methodology in maximizing the data rate, we can reformulate the SI power minimization problem as follows:

\vspace{-1.7em}
\begin{small}\begin{align}
 \mathop {\min }\limits_{\boldsymbol{\alpha}_e,\boldsymbol{\beta}_e}&\ \   {{\boldsymbol{\alpha }}_e^H}{\boldsymbol{\Xi} _{e,2}}{\boldsymbol{\alpha }_e} + {\mathbf{z}}_{e,2}^H{{\boldsymbol{\alpha }}_e^ * } + {{\boldsymbol{\alpha }}^T}{{\mathbf{z}}_{e,2}} \label{P21_OF}\\
{\text{s.t.}}  & \ \   {\text{diag}}\left\{ {{\boldsymbol{\alpha }_e}{{\boldsymbol{\alpha }}_e^H} + {\boldsymbol{\beta }_e}{{\boldsymbol{\beta }}_e^H}} \right\} \leq {{\mathbf{1}}_L},  \tag{\ref{P21_OF}{a}}  \label{P21_Power} \\
& \ \  2\operatorname{Re} \left\{ {{{\boldsymbol{\beta }}_e^H}\! {\boldsymbol{\Xi} _{e,1}}\! {\boldsymbol{\tilde \beta }_e}} \right\} \geq {{{\boldsymbol{\tilde \beta }}}_e^H}{\boldsymbol{\Xi} _{e,1}}{\boldsymbol{\tilde \beta }_e}\! \!  +\!\!  \left( {{2^{{R_{th}}}} \! -\!  1} \right)\! \sigma _d^2, \tag{\ref{P21_OF}{b}}  \label{P21_sinr}
\end{align}\end{small}which completes the joint amplitudes and phase shifts optimization part for ES-IOS. Note that the algorithm for minimizing the SI power can be found in Algorithm 3.
\\ \\
b) \textit{Optimizing reflecting and refracting phase shifts $\mathbf{\Theta '}_m$ and $\mathbf{\Phi '}_m$ for ES-IOS}

For the case with MS-IOS, following the similarity of the methodology above, the SI power minimization problem can be formulated as follows:

\vspace{-1.7em}
\begin{small}\begin{align}
 \mathop {\min }\limits_{\boldsymbol{\alpha}_m,\boldsymbol{\beta}_m}&\ \ {{\boldsymbol{\alpha }}_m^H}{\boldsymbol{\Xi} _{m,2}}{\boldsymbol{\alpha }_m} + {\mathbf{z}}_{m,2}^H{{\boldsymbol{\alpha }}_m^ * } + {{\boldsymbol{\alpha }}_m^T}{{\mathbf{z}}_{m,2}} + {d_{m,2}} \label{P22_OF}   \\
{\text{s.t.}} & \ \ 2\operatorname{Re} \left\{ {{{\boldsymbol{\beta }}_m^H}\! {\boldsymbol{\Xi} _{m,1}}\! {\boldsymbol{\tilde \beta }_m}} \right\} \geq {{{\boldsymbol{\tilde \beta }}}_m^H}{\boldsymbol{\Xi} _{m,1}}{\boldsymbol{\tilde \beta }_m}\! \!  +\!\!  \left( {{2^{{R_{th}}}} \! -\!  1} \right)\! \sigma _d^2, \tag{\ref{P22_OF}{a}}  \label{P22_rate}\\
& \ \  {\text{diag}}\left\{ {\boldsymbol{\alpha }_m}{{\boldsymbol{\alpha }}_m^H}  \right\} \leq {{\mathbf{1}}_L}, \tag{\ref{P22_OF}{b}} \label{P22_modulus1}\\
& \ \  {\text{diag}}\left\{ {\boldsymbol{\beta }_m}{{\boldsymbol{\beta }}_m^H}  \right\} \leq {{\mathbf{1}}_L}, \tag{\ref{P22_OF}{c}} \label{P22_modulus2}
\end{align}\end{small}which is a quadratically constrained quadratic problem and can be solved by CVX using QCQP method.

\begin{algorithm}[t]
\caption{SI power minimizations for both ES-IOS and MS-IOS}
\label{Minimizing}
\algsetup{linenosize=\footnotesize}
\footnotesize
\begin{algorithmic}
\STATE {Initialize the beamforming vector ${\mathbf{w}}^{0}$, the reflecting, refracting amplitudes and phase shifts ${\mathbf{\Theta}_e ^{0}}$, ${\mathbf{\Phi}_e ^{0}}$ of the ES-IOS,  the phase shifts  of MS-IOS  ${\mathbf{\Theta}_m ^{0}}$, ${\mathbf{\Phi}_m ^{0}}$, as well as the mode selection matrix of MS-IOS $\mathbf{A}_m^0$,
compute $P_e\left( {{\mathbf{w}}^0,\!{{\mathbf{\Theta }}_e^0},\!{{\mathbf{\Phi }}_e^0}} \right)$ and $P_m\left( {{\mathbf{w}}^0,\!{{\mathbf{\Theta }}_m^0},\!{{\mathbf{\Phi }}_m^0},{\mathbf{A}_m^0}} \right)$, respectively. Set $t=0$ and the accuracy for iteration $\varepsilon_2$.}
\REPEAT
\STATE  1. Given ${\mathbf{w}}^{ t }$, $\mathbf{A}_m^0$, ${\mathbf{\Theta}_e^{t}}$,  ${\mathbf{\Phi}_e^{t}}$, ${\mathbf{\Theta}_m^{t}}$ and ${\mathbf{\Phi}_m^{t}}$, compute the SI powers  as $P_{e}\left( {{\mathbf{w}}^{t}, {\mathbf{\Theta}_e ^{ t}}, {\mathbf{\Phi}_e ^{ t}}} \right)$ and $P_{m}\left( {{\mathbf{w}}^{t}, {\mathbf{\Theta}_m ^{ t}}, {\mathbf{\Phi}_m^{ t}}},{\mathbf{A}_m^t} \right)$, respectively.\\
2. Given ${\mathbf{\Theta}_e^{t}}$, ${\mathbf{\Phi}_e ^{t}}$ for the case with ES-IOS and given $\mathbf{A}_m^t$, ${\mathbf{\Theta}_m^{t}}$ and ${\mathbf{\Phi}_m ^{t}}$ for the case with MS-IOS, optimize ${\mathbf{w}}^{t+1}$  by solving problem \eqref{P19_OF}.\\
3. Given ${\mathbf{w}}^{t+1 }$, optimize ${\mathbf{\Theta} _e^{ t+1 }}$ and ${\mathbf{\Phi} _e^{ t+1 }}$ by solving problem \eqref{P21_OF} for the ES-IOS. As for the MS-IOS, optimize ${\mathbf{\Theta} _m^{ t+1 }}$ and ${\mathbf{\Phi} _m^{ t+1 }}$ with given $\mathbf{w}^{t+1}$ and $\mathbf{A}_m^t$ for the case with MS-IOS, by solving problem \eqref{P22_OF}. \\
4. By solving \eqref{P23_OF} for the case with MS-IOS, optimize $\mathbf{X}$ given $\mathbf{w}^{t+1}$, $\mathbf{\Theta}_m^{t+1}$ and $\mathbf{\Phi}_m^{t+1}$.\\
5. For the case with MS-IOS, retrieve $\mathbf{A}_m^{t+1}$ from $\mathbf{X}$ using Gaussian randomization procedure. \\
6. Compute the SI powers $P_{e}\left( {{\mathbf{w}}^{t+1}, {{\mathbf{\Theta }}_e^{t+1}},{{\mathbf{\Phi}}_e^{t+1}}} \right)$ and $P_{m}\left( {{\mathbf{w}}^{t+1}, {{\mathbf{\Theta }}_m^{t+1}},{{\mathbf{\Phi}}_m^{t+1}},\mathbf{A}_m^t} \right)$ for both the IOSs.\\
7. Set $t = t+1$.
\UNTIL {$\frac{{\left| {{P_{e}}\left( {{\mathbf{w}}^{t + 1},{\mathbf{\Theta }}_e^{t + 1},{\mathbf{\Phi }}_e^{t + 1}} \right) - {P_{e}}\left( {{\mathbf{w}}^t,{{\mathbf{\Theta }}_e^t},{{\mathbf{\Phi }}_e^t}} \right)} \right|}}{{{P_{e}}\left( {{\mathbf{w}}^{t + 1}, {{\mathbf{\Theta }}_e^{t + 1}},{{\mathbf{\Phi }}_e^{t + 1}}} \right)}} \leq \varepsilon_2$  or  $\frac{{\left| {{P_{m}}\left( {{\mathbf{w}}^{t + 1},{\mathbf{\Theta }}_m^{t + 1},{\mathbf{\Phi }}_m^{t + 1}, \mathbf{A}_m^{t+1}} \right) - {P_{m}}\left( {{\mathbf{w}}^t,{{\mathbf{\Theta }}_m^t},{{\mathbf{\Phi }}_m^t},\mathbf{A}_m^t} \right)} \right|}}{{{P_{m}}\left( {{\mathbf{w}}^{t + 1},{{\mathbf{\Theta }}_m^{t + 1}},{{\mathbf{\Phi }}_m^{t + 1}},\mathbf{A}_m^{t+1}} \right)}} \leq \varepsilon_2$.}
\end{algorithmic}
\end{algorithm}

\subsection{Optimizing mode selection $\mathbf{A}_m$ with given $\mathbf{w}$, $\mathbf{\Theta '}_m$ and $\mathbf{\Phi '}_m$ for MS-IOS}
In this subsection, we aim to optimize the mode selection matrix $\mathbf{A}_o$ with given phase shifts and beamforming vector for the case with MS-IOS. The SI power minimization problem is formulated as follows:

\vspace{-1.7em}
\begin{small}\begin{align}
 \mathop {\min }\limits_{\mathbf{X}}&\ \ \frac{1}{4}{\text{Tr}}\left( {{{{\mathbf{\Xi '}}}_{m,2}}{\mathbf{X}}} \right) + {c_{m,2}}  \label{P23_OF}   \\
{\text{s.t.}} & \ \ \frac{1}{4}{\text{Tr}}\left( {{{{\mathbf{\Xi '}}}_{m,1}}{\mathbf{X}}} \right) + {c_{m,1}} \geq\!  (2^{R_{th}}-1)\sigma _d^2, \tag{\ref{P23_OF}{a}}  \label{P23_sinr}\\
& \ \ {\text{diag}}\left\{ {\mathbf{X}} \right\} = {{\mathbf{1}}_{L + 1}}, \tag{\ref{P23_OF}{b}} \label{P23_1L}
\end{align}\end{small}which is a convex problem with convex constraints, and
shares the similarity as the data rate maximization problem which can be easily solved using the SDR method.

\subsection{Overall Algorithm and  complexity analysis}
To sum up, the overall alternative optimization algorithm for minimizing the SI is provided in Algorithm 3.  The QCQP method is applied to solve the beamforming vector, reflecting, reflecting joint amplitudes and  phase shifts optimizations for the case with ES-IOS, as well as the phase shifts optimization for the MS-IOS.  Furthermore, the SDR method and Gaussian randomization procedure are exploited for solving the element mode selection for the case with MS-IOS. Every step in Algorithm 3 guarantees the objective function decreases monotonically, therefore, the convergence of the proposed algorithm is guaranteed.  The overall computational complexities of SI minimization problem for the cases with ES-IOS and MS-IOS are $\mathcal{O}\left( {{N_{ite}} \times \max \left\{ {M^3,{L^3}} \right\}} \right)$ and $\mathcal{O}\left( {{N_{ite}} \times \max \left\{ {M^3,{L^{3.5}},G} \right\}} \right)$, respectively.

\section{Numerical Results}
In this section, simulation results are provided to demonstrate the performance of the proposed optimization methods. The 3-dimensional coordinates of the destination, the first transmit antenna, the first receive antenna, and the first element at the IOS are set as $\left[ {20, - 10,1.5} \right]$, $\left[ {0,0,5} \right]$, $\left[ {0,0.1,5} \right]$ and $\left[ {0.5,0,5} \right]$ m, respectively.  In addition, $\lambda=0.05$ m and the distance between any two adjacent elements/antennas is $\frac{\lambda }{2}$,  the Rician path-loss exponent and Rician factor are  $\kappa=2.5$ and $K=3 $ dB, respectively. The noise powers at the destination and transmitter are $n_d=n_r=-80$ dBm,  the SI power threshold $P_{th}$ and the data rate threshold for the optimizations are $-90 \sim -10$ dBm and $0.5 \sim 3$ bps/Hz, respectively. Furthermore, the length of the Gaussian randominization procedure $G=10^{3}$, and the convergence thresholds are set to $\varepsilon_1  =\varepsilon_2= {10^{ - 5}}$.  Specifically, we compare the performance of our proposed algorithms with the following three schemes:
\begin{enumerate}
\item \textbf{WO-IOS}: In this scheme, only the beamforming vectors of the transmit signals are optimized in the proposed system without the assistance of IOS. This scheme is set as the benchmark for performance evaluation.

\item \textbf{ES-IOS}: The ES-IOS is employed to assist signal transmission and SI mitigation; specifically, the beamforming vectors of the transmitter, as well as the reflecting and refracting phase shifts of the ES-IOS, are designed using the QCQP method.

\item \textbf{MS-IOS}: Compared with the ES-IOS, each element of the MS-IOS can either reflect or refract signals in one time slot; therefore, the beamforming vectors, both the reflecting and the refracting phase shifts, jointly with the mode selection, are optimized for the proposed system.
\end{enumerate}

\begin{figure}[!t]
        \centering
        \includegraphics*[width=90mm]{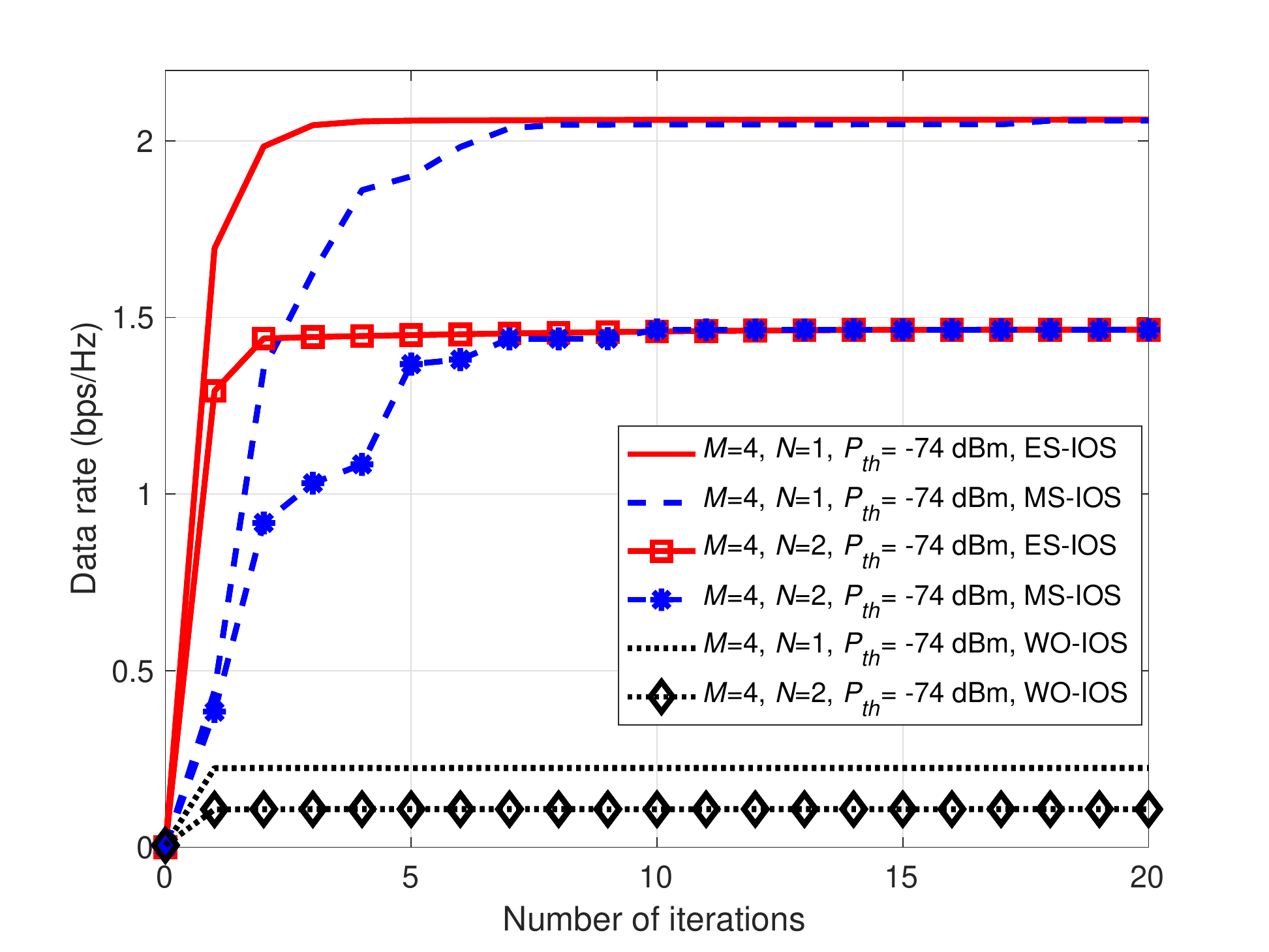}
       \caption{Iteration behavior of the data rate maximization problem with the three schemes.}
        \label{fig2}
\end{figure}

\begin{figure}[!t]
        \centering
        \includegraphics*[width=90mm]{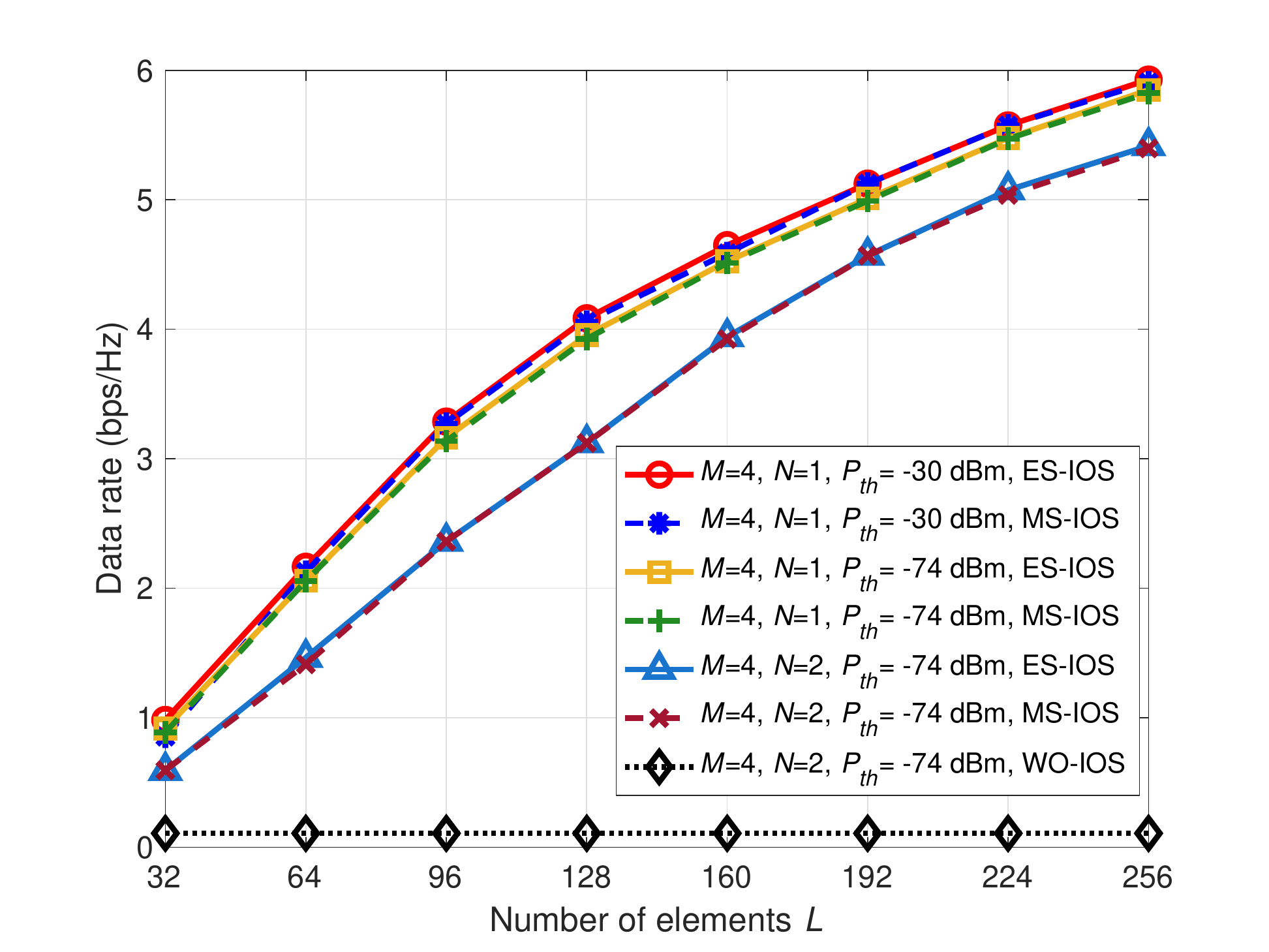}
       \caption{Data rate performance of the three schemes with different $L$ and $N$.}
        \label{fig3}
\end{figure}

Fig. \ref{fig2} shows the data rate convergence behaviours of the proposed joint optimization method with SI power threshold $P_{th}=-74$ dBm, and different numbers of elements $L$ and numbers of receive antenna $N$ for the three schemes. It is shown that the data rates of all schemes converge fast with the proposed algorithms, which demonstrates their efficiency.  Furthermore, the data rates decrease with $N$, and it is evident that both ES-IOS and MS-IOS outperform WO-IOS in performance due to the degrees of freedom bestowed by the IOSs. Besides, the data rates with MS-IOS are slightly lower than that of ES-IOS because each element of the MS-IOS can only reflect or refract signals at one time, and it can not fully exploit the reflecting and refracting resource as ES-IOS does.

The comparison of data rates with different numbers of elements $L$ and different numbers of receive antennas $N$ is illustrated in Fig. \ref{fig3}. As we can see, to guarantee the SI power below a threshold $P_{th}=-74$ dBm, the case with WO-IOS can not achieve a high data rate. In addition, it can be noticed that the data rates with ES-IOS and MS-IOS increase with the number of elements, and the performance of ES-IOS is close to that of MS-IOS. Significant improvement can be achieved by introducing ES-IOS or MS-IOS into FD wireless communication networks; for example,  $5.85$ and $5.82$ bps/Hz can be obtained with ES-IOS and MS-IOS, respectively, when $L=256$, $M=4$ and $N=1$. However, without the assistance of an IOS, only the beamforming can be designed to mitigate the SI; thus, only $0.22$ bps/Hz can be achieved. In addition, the data rates with $P_{th}=-30$ dBm are larger compared with  $P_{th}=-74$ dBm due to the lower SI mitigation requirement, e.g. $4.08$ bps/Hz with $L=128$, $M=4$, $N=1$ for ES-IOS case, around $4\%$ larger than the latter case.

\begin{figure}[!t]
        \centering
        \includegraphics*[width=90mm]{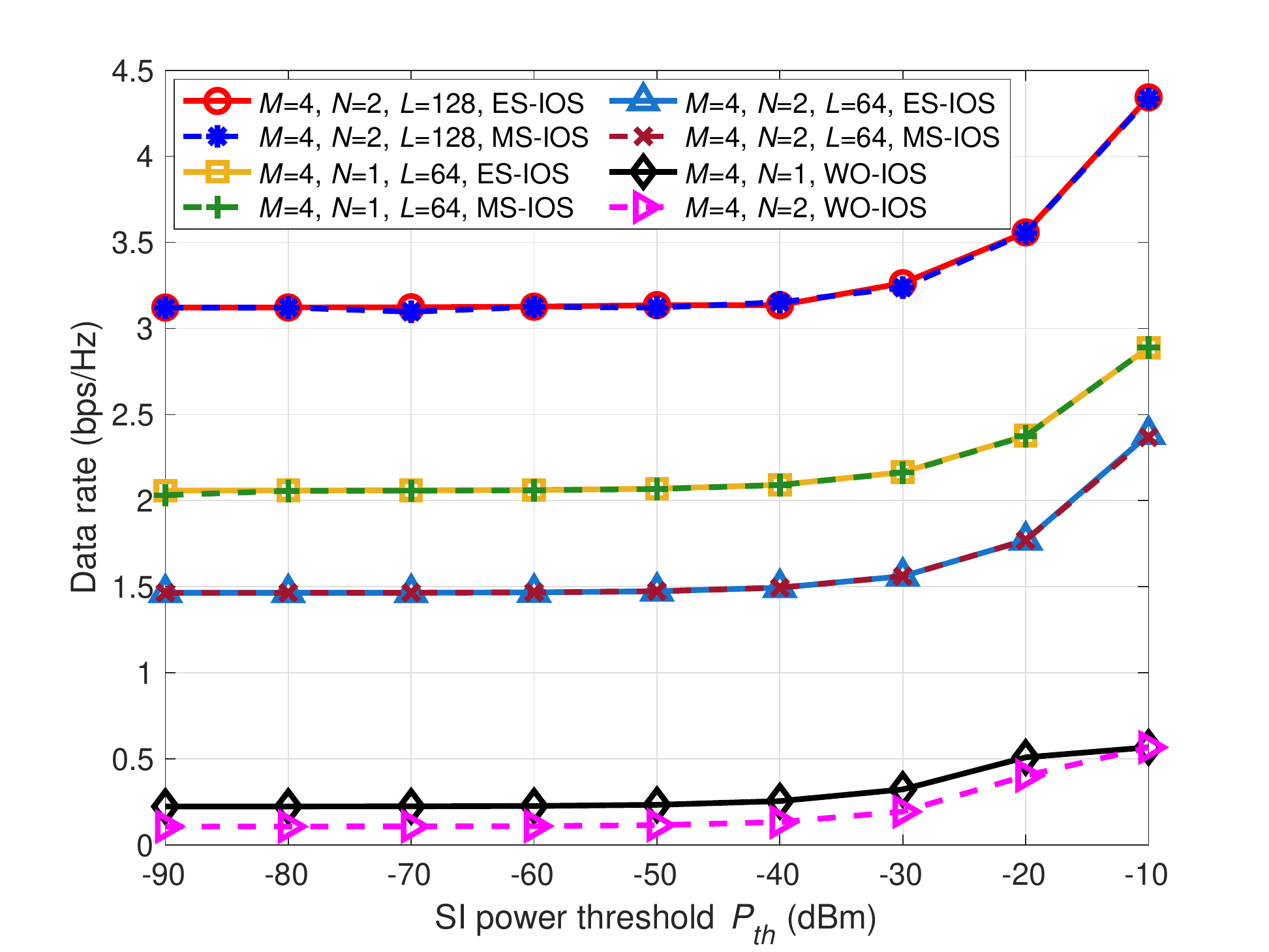}
       \caption{Data rate performance of the three schemes with different SI threshold $P_{th}$.}
        \label{fig4}
\end{figure}

Fig. \ref{fig4}  illustrates the data rate performance of the proposed algorithm with $L=64/128$ and different SI power thresholds for the three schemes. It can be noticed that the data rate grows slowly with the SI power threshold (from $-90$ dBm to $-40$ dBm) for ES-IOS and MS-IOS, which implies that both IOSs are capable of reducing SI at the cost of negligible data rate degradation. Also, if we reduce the requirement of the SI thresholds from $-40$ dBm to $-10$ dBm, the transmission rate will be increased significantly. Furthermore, the cases without an IOS are introduced for comparison; both data rates with $N=2$ and $N=1$ are much lower than that with ES-IOS and MS-IOS.
\begin{figure}[!t]
        \centering
        \includegraphics*[width=90mm]{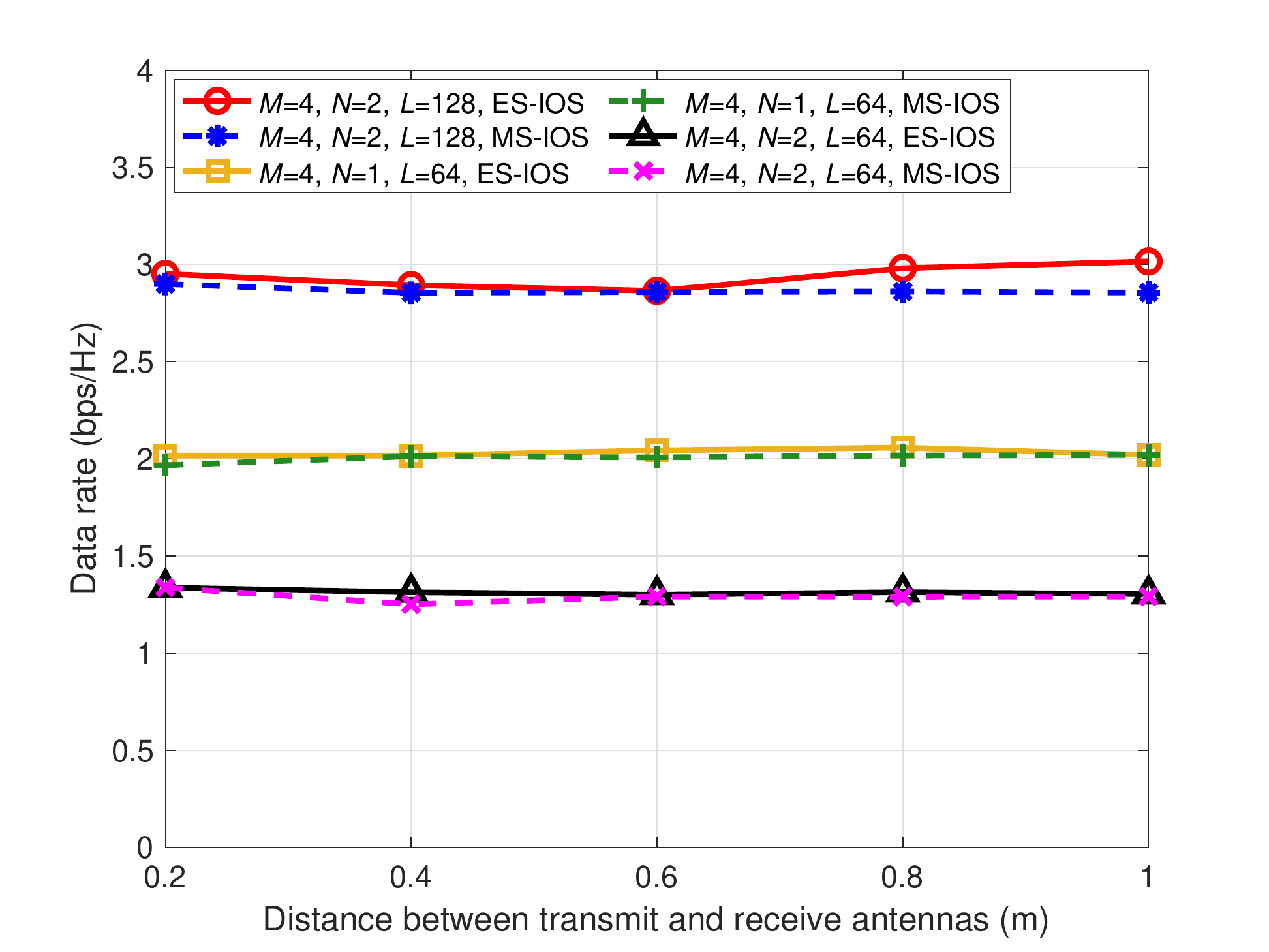}
       \caption{The impact of the distance between transmit and receive antennas on the data rates.}
        \label{fig5}
\end{figure}
\begin{figure}[!t]
        \centering
        \includegraphics*[width=90mm]{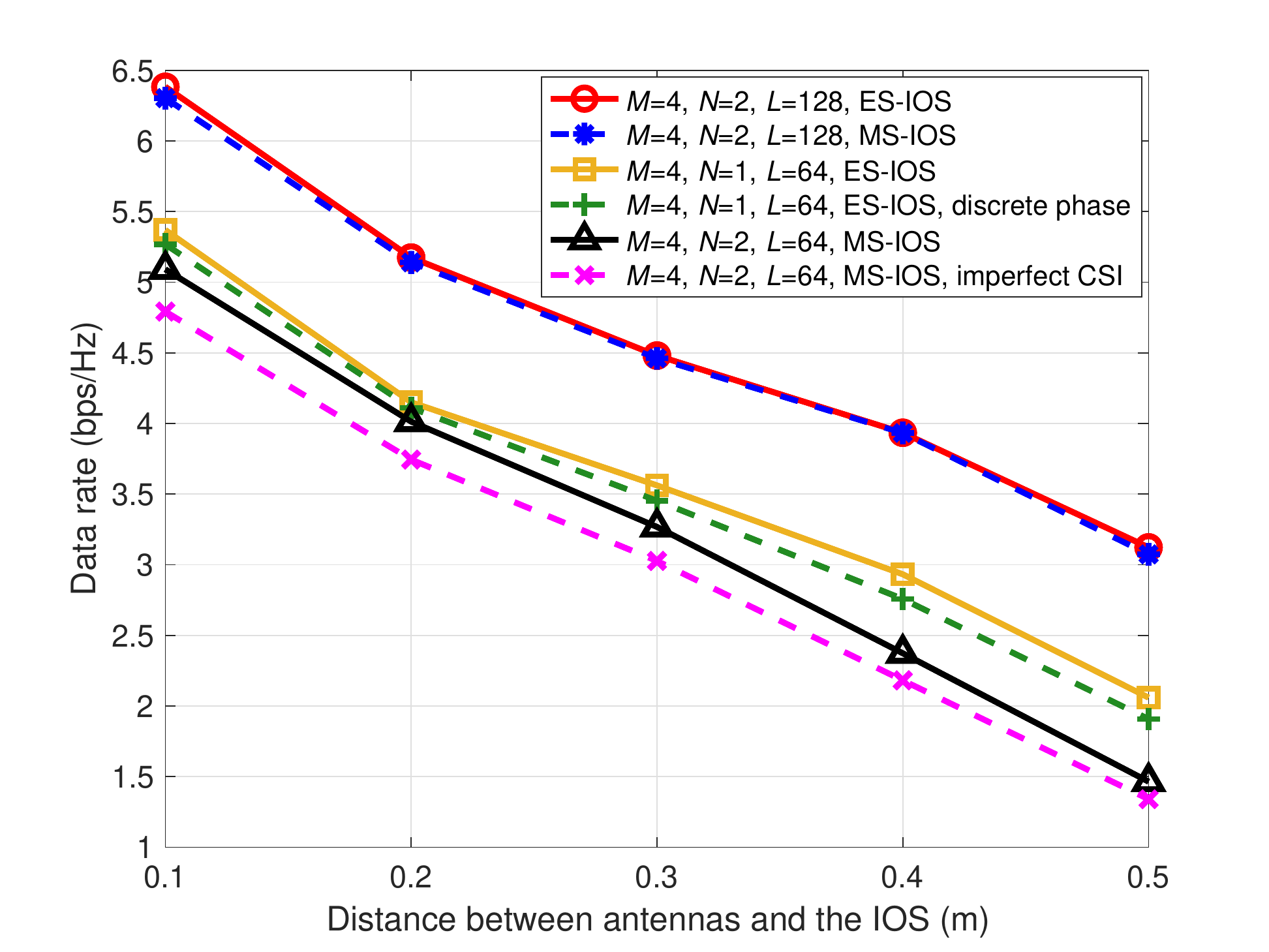}
       \caption{The impact of the distance between  transmitter and IOSs on the data rates.}
        \label{fig6}
\end{figure}

In addition, the impact of the distances between transmit and receive antennas on the data rate is shown in Fig. \ref{fig5}. It is shown that the data rates of the proposed schemes remain stable for the distance from $0.2$ m to $1$ m with different $L$ and $N$, because multiple transmit and receiver antennas and the IOS are considered in the proposed system. Using the IOS will guarantee transmission and reflection beam patterns are the dominant factors to balance the tradeoff between the date rate and SIC against the distance's effect.  Also, it is evident that for all curves, the data rates with the ES-IOS case are slightly larger than that with the MS-IOS case.

Fig. \ref{fig6} illustrates the impact of the distance between  transmitter and IOSs on the data rates. The curves clearly show that the data rates decrease with the distance between transmitter and IOSs. Furthermore, the data rates with continuous phase shifts slightly outperforms the discrete phase shifts with 4 bits quantization with $M=4$, $N=1$ and $L=64$ for the ES-IOS case, e.g. $2.06$ bps/Hz with continuous phase shifts compared to $1.91$ bps/Hz with discrete phase shifts.  Also, for the case of imperfect channel estimation, we define the estimated channels between the IOSs and destination, the transmit and receive antennas as ${{{\mathbf{\bar h}}}_{id}} = \sqrt \eta  {{\mathbf{h}}_{id}} + \sqrt {1 - \eta } \Delta {{\mathbf{h}}_{id}}$ and ${{{\mathbf{\bar H}}}_{tr}} \!=\! \sqrt \eta  {{\mathbf{H}}_{tr}} \!+\! \sqrt {1 \!-\! \eta } \Delta {{\mathbf{H}}_{tr}}$, where $\eta \! \in\! \left[ {0,1} \right]$ represents the estimation accuracy, $\Delta {{\mathbf{h}}_{id}}$ and $\Delta {{\mathbf{H}}_{tr}}$ are the estimation errors, which have the same statistical properties as of ${{\mathbf{h}}_{id}}$ and ${{\mathbf{H}}_{tr}}$ in terms of mean and variance values. The rate performance loss can be observed when imperfect CSI is considered for MS-IOS case with $M=4$, $N=2$ and $\eta=0.95$, for example, $1.34$ bps/Hz compared to $1.47$ bps/Hz with perfect CSI when the distance between the transmitter and the IOS is $0.5$ m.

\begin{figure}[!t]
        \centering
        \includegraphics*[width=90mm]{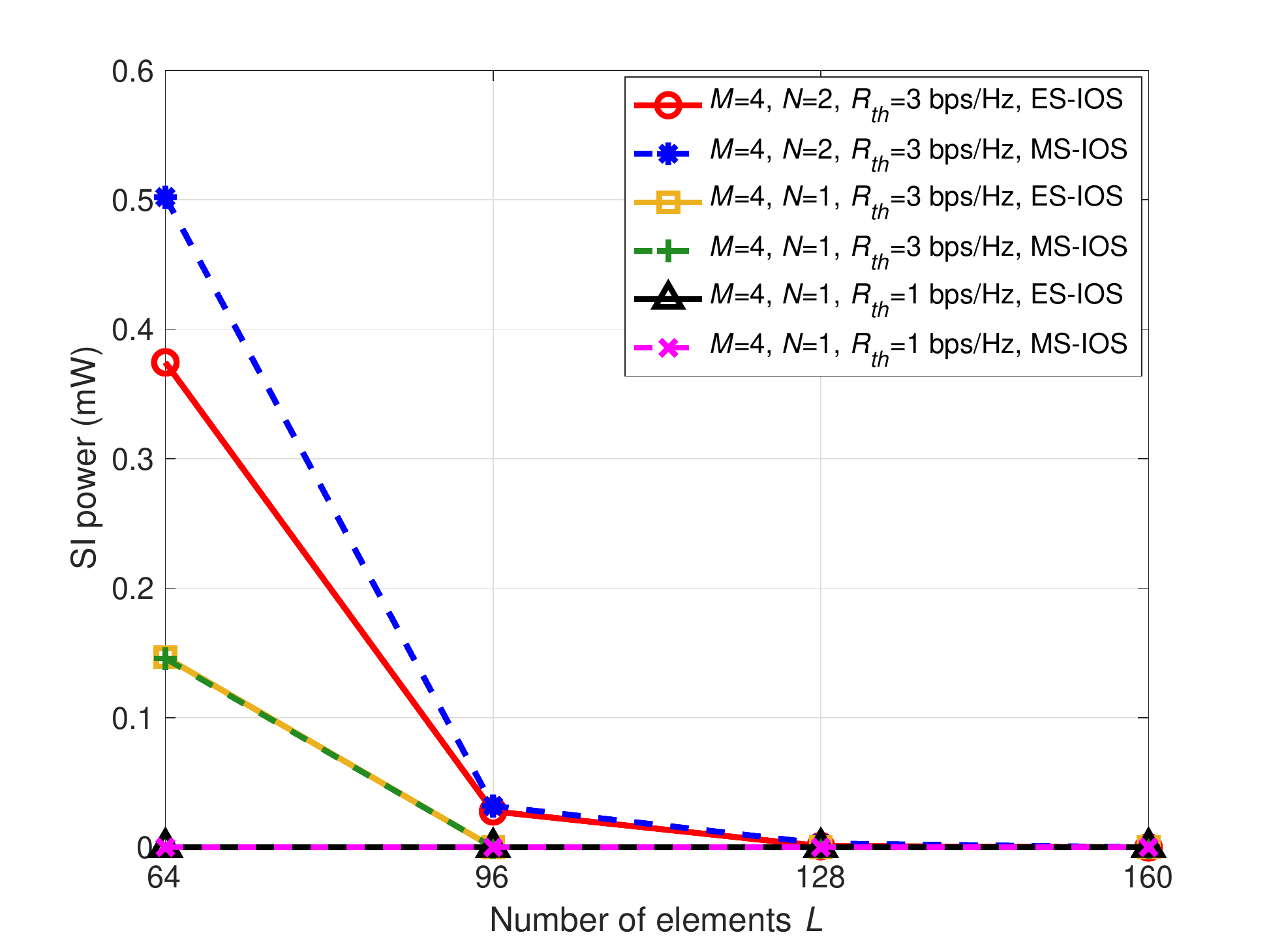}
       \caption{SI power performance of the three schemes with different $L$ and $N$.}
        \label{fig7}
\end{figure}

\begin{figure}[!t]
        \centering
        \includegraphics*[width=90mm]{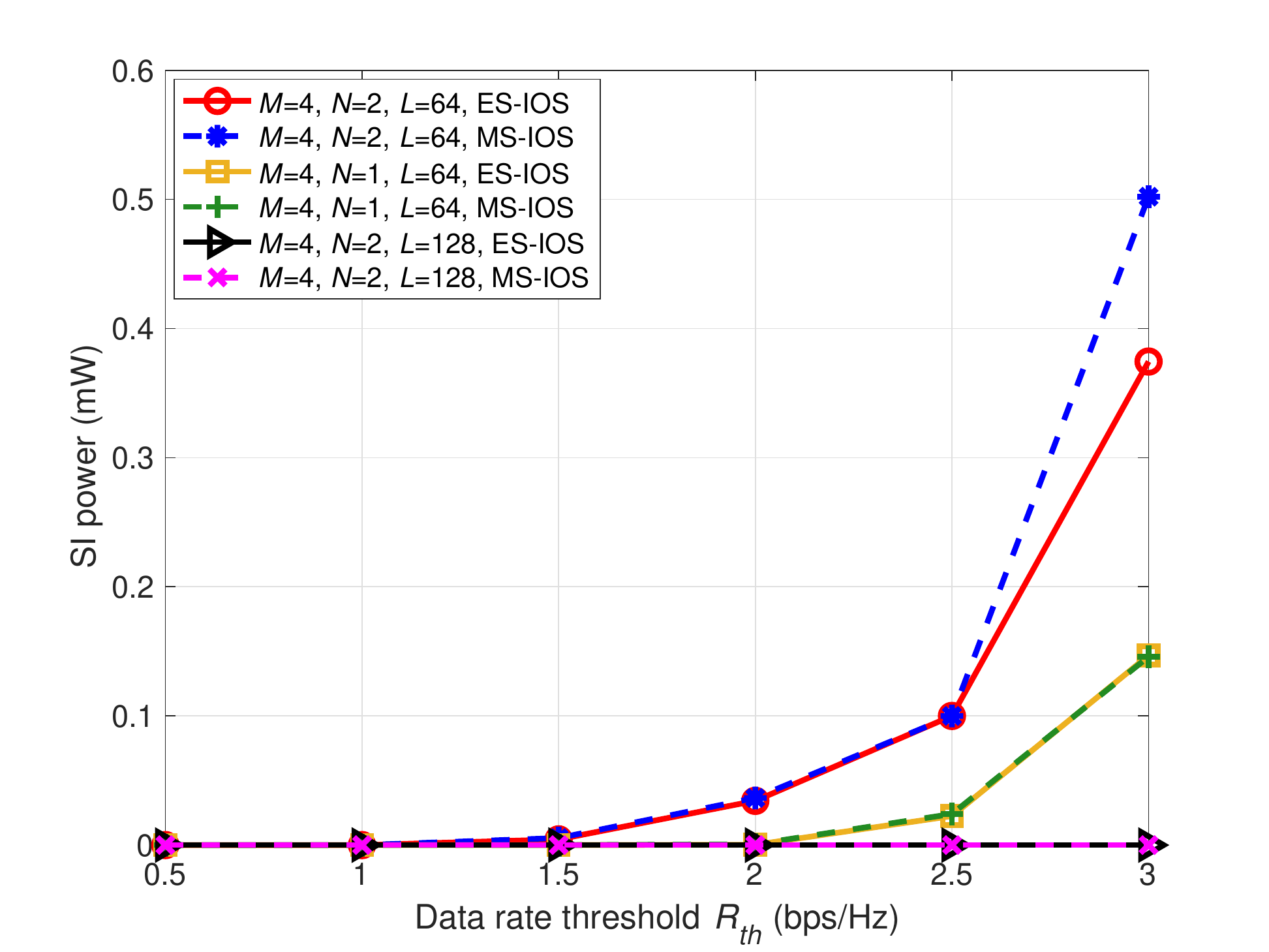}
       \caption{SI power performance of the three schemes with different data rate threshold $R_{th}$.}
        \label{fig8}
\end{figure}

The performance of the SI power minimization concerning different numbers of receive antennas $N$ and different numbers of elements $L$ is shown in Fig. \ref{fig7}. The data rate threshold $R_{th}=1$ and $3$ bps/Hz are introduced for both ES-IOS and MS-IOS.  It is obvious that the SI powers of ES-IOS and MS-IOS reach nearly zero with $L \geq 96$ for $R_{th}=1$ bps/Hz; and with $L \geq 128$ for $R_{th}=3$ bps/Hz.  However, the SI powers of $L<96$ with $R_{th}=3$ bps/Hz is still large for both ES-IOS and MS-IOS, the reason for it is to meet a certain rate threshold with small $L$, a large proportion of the IOS elements should fully participate in increasing the rate, thus renders high SI.   Furthermore, the SI powers with MS-IOS are larger than that with ES-IOS, e.g.  $0.032$ mW for MS-IOS compared to $0.028$ mW for ES-IOS with $M=4$, $N=2$, $L=96$ and $R_{th}=3$ bps/Hz. In addition, it is evident that SI powers with $R_{th}=1$ bps/Hz for both IOSs are smaller than that with $R_{th}=3$ bps/Hz, and it can be noticed that fewer receive antennas lead to lower SI powers. Nevertheless, even with a high rate requirement, both IOSs are still capable of achieving the required data rate, as well as reducing the SI power to nearly zero, via deploying more reflecting and refracting elements.

Fig. \ref{fig8}  shows the SI powers with different $L$ and different rate thresholds for the cases with ES-IOS and MS-IOS. It can be observed that a higher data rate requirement would lead to higher residual SI power, and the SI mitigation performance with ES-IOS is slightly better than that with MS-IOS. Also, it is evident that the numbers of receive antennas impact the SI power significantly in the proposed FD wireless communication system.

\section{Conclusion}
In this paper, we proposed a new IOS-assisted FD MISO implementation, which solved the frequency-dependent and SWaP limitation in the traditional FD scheme by completely bypassing the analogue SIC. To maximize the data rate at the destination and minimize SI power received at the antennas, beamforming vectors at the transmitter, the phase shifts, and mode selection optimizations of ES-IOS and MS-IOS were conducted. However, both the formulated problems were non-convex and intractable due to the coupling of the variables. Alternative optimization algorithms were designed to circumvent this issue to solve the problems iteratively. Specifically, the QCQP method was applied to solve the beamforming vectors and the phase shifts for ES-IOS cases. On the other hand, the mode selection optimization for the case with MS-IOS was still intractable. Hence, we resorted to the SDR and Gaussian randomization procedure to solve it. Furthermore, the first-order Taylor expansions were leveraged to convexify the non-convex constraints for the optimization problems.  The simulation results demonstrated the efficacy of the proposed algorithms and the flexibility of utilizing the IOSs for trading off the SIC and data transmission.

\begin{appendices}
\section{Derivations of (37) and (38)}
With the binary variable ${\mathbf{b}} = 2{\mathbf{a}_m} - {{\mathbf{1}}_L}$, the objective function of the data rate maximization problem can be reformulated as follows:
\begin{equation}\small
\begin{gathered}
  {\mathbf{a}}_m^T{{\mathbf{\Xi }}_{m,1}}{{\mathbf{a}}_m} - 2\operatorname{Re} \left\{ {{\mathbf{a}}_m^T{{\mathbf{w}}_{m,1}}} \right\} + {d_{m,1}} \hfill \\
   = {\mathbf{a}}_m^T{{\mathbf{\Xi }}_{m,1}}{{\mathbf{a}}_m} + 2\operatorname{Re} \left\{ {{\mathbf{a}}_m^T\left( { - {{\mathbf{w}}_{m,1}}} \right)} \right\} + {d_{m,1}} \hfill \\
   = {\left( {\frac{{{{\mathbf{1}}_L} + {\mathbf{b}}}}{2}} \right)^T}{{\mathbf{\Xi }}_{m,1}}\left( {\frac{{{{\mathbf{1}}_L} + {\mathbf{b}}}}{2}} \right) + {d_{m,1}} + 2\operatorname{Re} \left\{ {{{\left( {\frac{{{{\mathbf{1}}_L} + {\mathbf{b}}}}{2}} \right)}^T}\left( { - {{\mathbf{w}}_{m,1}}} \right)} \right\} \hfill \\
   = \frac{1}{4}{\text{Tr}}\left( {{{\mathbf{\Xi }}_{m,1}}{\mathbf{B}}} \right) + \frac{1}{4}{\text{Tr}}\left( {{{\mathbf{\Xi }}_{m,1}}} \right) + \frac{1}{2}{\text{Tr}}\left( {{{\mathbf{\Xi }}_{m,1}}{{\mathbf{1}}_L}{{\mathbf{b}}^T}} \right) \hfill \\
   \quad\  + \operatorname{Re} \left\{ {{{\mathbf{b}}^T}\left( { - {{\mathbf{w}}_{m,1}}} \right)} \right\} + \operatorname{Re} \left\{ {{\mathbf{1}}_L^T\left( { - {{\mathbf{w}}_{m,1}}} \right)} \right\} + {d_{m,1}} \hfill \\
   = \frac{1}{4}\left( {{\text{Tr}}\left( {{{\mathbf{\Xi }}_{m,1}}{\mathbf{B}}} \right) + 2\operatorname{Re} \left\{ {{{\mathbf{h}}^T}{\mathbf{b}}} \right\}} \right) + {c_{m,1}}, \hfill \\
\end{gathered}
\end{equation}
where ${\mathbf{B}} = {\mathbf{b}}{{\mathbf{b}}^T}$, ${\mathbf{h}} =  - 2{{\mathbf{w}}_{m,1}} + {{\mathbf{h}}_{m,1}}$, and
\begin{equation}\small
{{\mathbf{h}}_{m,1}} = {\left[ {\sum\limits_{i = 1}^L {{{\left[ {{{\mathbf{\Xi }}_{m,1}}} \right]}_{1,i}}} ;\sum\limits_{i = 1}^L {{{\left[ {{{\mathbf{\Xi }}_{m,1}}} \right]}_{2,i}}} ; \ldots ;\sum\limits_{i = 1}^L {{{\left[ {{{\mathbf{\Xi }}_{m,1}}} \right]}_{L,i}}} } \right]^T},
\end{equation}
\begin{equation}\small
{c_{m,1}} = \frac{1}{4}{\text{Tr}}\left( {{{\mathbf{\Xi }}_{m,1}}} \right) + \operatorname{Re} \left\{ {{\mathbf{1}}_L^T\left( { - {{\mathbf{w}}_{m,1}}} \right)} \right\} + {d_{m,1}}.
\end{equation}
Similarly, we can obtain the SI power function as follows:
\begin{equation}\small
\frac{1}{4}\left( {{\text{Tr}}\left( {{{\mathbf{\Xi }}_{m,2}}{\mathbf{B}}} \right) + 2\operatorname{Re} \left\{ {{{\mathbf{g}}^T}{\mathbf{b}}} \right\}} \right) + {c_{m,2}},
\end{equation}
where
\begin{equation}\small
{\mathbf{g}} =  2{{\mathbf{w}}_{m,2}} + {{\mathbf{h}}_{m,2}},
\end{equation}
\begin{equation}\small
{{\mathbf{h}}_{m,2}} = {\left[ {\sum\limits_{i = 1}^L {{{\left[ {{{\mathbf{\Xi }}_{m,2}}} \right]}_{1,i}}} ;\sum\limits_{i = 1}^L {{{\left[ {{{\mathbf{\Xi }}_{m,2}}} \right]}_{2,i}}} ; \ldots ;\sum\limits_{i = 1}^L {{{\left[ {{{\mathbf{\Xi }}_{m,2}}} \right]}_{L,i}}} } \right]^T},
\end{equation}
\begin{equation}\small
{c_{m,2}} = \frac{1}{4}{\text{Tr}}\left( {{{\mathbf{\Xi }}_{m,2}}} \right) + \operatorname{Re} \left\{ {{\mathbf{1}}_L^T\left( { - {{\mathbf{w}}_{m,2}}} \right)} \right\} + {d_{m,2}}.
\end{equation}
Then by introducing another binary variable ${\mathbf{x}} = {\left[ {{\mathbf{b}};1} \right]^T}$, we can reformulate the objective function and SI power as follows:
\begin{equation}\small
\frac{1}{4}{\text{Tr}}\left( {{{{\mathbf{\Xi '}}}_{m,1}}{\mathbf{X}}} \right) + {c_{m,1}},
\end{equation}
\begin{equation}\small
\frac{1}{4}{\text{Tr}}\left( {{{{\mathbf{\Xi '}}}_{m,2}}{\mathbf{X}}} \right) + {c_{m,2}},
\end{equation}
respectively, where ${{{\mathbf{\Xi '}}}_{m,1}} = \left[ \begin{gathered}
  {{\mathbf{\Xi }}_{m,1}}\quad {\mathbf{h}} \hfill \\
  {{\mathbf{h}}^T}\quad\ \  \ 0 \hfill \\
\end{gathered}  \right]$, ${{{\mathbf{\Xi '}}}_{m,2}} = \left[ \begin{gathered}
  {{\mathbf{\Xi }}_{m,2}}\quad {\mathbf{g}} \hfill \\
  {{\mathbf{g}}^T}\quad\ \ \ 0 \hfill \\
\end{gathered}  \right]$ and ${\mathbf{X}} = {\mathbf{x}}{{\mathbf{x}}^T}$, which completes the derivation.
\end{appendices}

\footnotesize
\bibliographystyle{IEEEtran}
\bibliography{ref}
\end{document}